\documentclass[journal]{IEEEtran}

\usepackage{graphics, psfrag, epsfig, amsmath, cite, amssymb, amsbsy, amsthm, url, setspace, amsfonts, xcolor, longtable, array, caption, mathrsfs, multicol, graphicx, varioref, pifont, lettrine}
\usepackage[export]{adjustbox}

\usepackage{algorithm}
\usepackage{algpseudocode}
\usepackage{algorithmicx}

\newcommand{\EqVspace}{0cm}
\newcommand{\AfterSectionTitle}{0cm}
\newcommand{\BeforeSectionTitle}{0cm}
\newcommand{\CMTcolor}{teal}
\newcounter{MYtempeqncnt}

\title{
Queue-Aware Joint Dynamic Interference Coordination and Heterogeneous QoS Provisioning in OFDMA Networks}
\author{Alireza~Sharifian, \textit{Member, IEEE} and Raviraj Adve, \textit{Fellow, IEEE}
\thanks{
Manuscript received Nov. 17, 2016; revised June 9, 2017 and Nov. 26, 2017; accepted Jan. 9, 2018.
The authors are with the Dept. of Elec.~and Comp.~Eng., University of Toronto, ON., Canada, \{alirezash, rsadve\}@ece.utoronto.ca.
This work was supported by TELUS Canada and Ontario Centres of Excellence (OCE). } }

\begin{document}
\date{}
\maketitle
\pagestyle{plain}
\vspace{-0.5cm}
\begin{abstract}
We propose algorithms for cloud radio access networks that not only provide heterogeneous quality of-service (QoS) for rate- and, importantly, delay-sensitive applications, but also jointly optimize the
frequency reuse pattern. Importantly, unlike related works, we account for random arrivals, through
queue awareness and, unlike majority of works focusing on a single frame only, we consider QoS measures averaged over multiple frames involving a set of closed loop controls. We model this problem as multi-cell optimization to maximize a sum utility subject to the QoS constraints, expressed as minimum mean-rate or maximum mean-delay. Since we consider dynamic interference coordination jointly with dynamic user association, the problem is not convex, even after integer relaxation. We translate the problem into an optimization of frame rates, amenable to a decomposition into intertwined primal and dual problems. The solution to this optimization problem provides joint decisions on scheduling, dynamic interference coordination, and, importantly, unlike most works in this area, on dynamic user association. Additionally, we propose a novel method to manage infeasible loads. Extensive simulations confirm that the design responds to instantaneous loads, heterogeneous user and AP locations, channel conditions, and QoS constraints while, if required, keeping outage low when dealing with infeasible loads. Comparisons to the baseline proportional fair scheme illustrate the gains achieved.
\end{abstract}


\vspace{-\BeforeSectionTitle}
\section{Introduction}
\vspace{-\AfterSectionTitle}
\lettrine[findent=1pt]{{\textbf{U}}}{}
biquitous connectivity is a key goal in designing wireless networks enabling broad ranges of reliable services to users. Cellular networks are evolving toward a distributed access point (AP) architecture controlled remotely over a cloud radio access network (C-RAN). By coordinating transmissions across APs, the C-RAN approach provides many benefits including cost, coverage, and capacity improvements. The resulting architecture, with edgeless virtual cells, meeting heterogeneous quality-of-service (QoS) metrics, is crucial to future wireless networks~\cite{3gppIMT2020}.

Designing for heterogeneity in user demands is relatively new; historically, traffic has been assumed homogeneous in time and space, therefore, interference coordination was performed by static frequency planning. More recently, reuse-1 (reusing frequency resources potentially everywhere) enhances throughput, but largely ignores users' QoS demands. At best, LTE schedulers are allowed to identify applications as guaranteed bit rate (GBR) or non-GBR. However, networks now deal with a broad range of applications, some that are delay-sensitive (DS), some rate-sensitive (RS), and others that just require best effort (BE). These complex \emph{heterogeneous} demands cannot be served effectively without advanced adaptation of dynamic interference coordination, dynamic user association, and fine-grained scheduling.

The growing heterogeneity in applications, and in traffic distributions in space and time, motivates changing the network architecture from assigning resources a-priori to APs, toward assigning resources dynamically to the users. In such a design, a user may be associated with multiple APs and the association may change over time, based on both \emph{channel} and \emph{AP load} conditions. This flexibility helps meet QoS constraints and allows for offloading to under-utilized cells making the frequency resources to \textit{follow} the traffic loads and be reused adaptively.

In this paper, we consider radio resource management (RRM), in a multiuser orthogonal frequency division multiple access (OFDMA) network. Unlike other works, we consider \textit{dynamic} interference and \textit{dynamic} user association (also called short-term user association), while jointly addressing \textit{heterogeneous multiple QoS}. Crucially, our \textit{finite backlog, queue-aware} formulation, with \textit{random arrivals} addresses \textit{delay sensitive} flows. Importantly, unlike related works, this allows us to \textit{avoid} treating a delay constraint as equivalent to a constant rate guarantee. Moreover, unlike many other works~\cite{Tao2008, Zhang2014, Zhang2015, Zhang2016}, we no not use time-sharing and will provide explicit scheduling. We also provide solutions to manage infeasible load conditions, when the core optimization problem becomes infeasible (due to the high input load) making the setup robust to the input load.

Delay is a measurement across frames, i.e., inherently we have a multi-frame problem. This is in contrast to many other works focusing on single frames. Therefore, we first translate the multi frame problem (through a set of closed loop controls) into an optimization of frame rates. Because of our assumptions of \textit{dynamic interference and dynamic user association}, our formulation results in a complex problem, preventing the use of conventional methods. We devise several techniques to develop an effective iterative QoS aware interference coordination (\textsc{QoSaIC}) algorithm for this challenging problem. We then propose a systematic approach for infeasible load conditions, combining the \textsc{QoSaIC} algorithm with an infeasible load management (\textsc{ILM}) algorithm.

\vspace{-\BeforeSectionTitle}
\section{Related Works, Research Gap, Approach, and Contributions}
\vspace{-\AfterSectionTitle}
\subsection{Related Works and Identifying the Research Gap}
\vspace{-\AfterSectionTitle}
\label{sec:RelatedWorks}
In this section, to identify the research gap addressed in this paper, we place the related literature into four categories. As a summary, the first category focuses on the QoS without interference awareness; the second one considers exclusively static interference coordination (with or without QoS awareness); the third set, addresses dynamic interference coordination but without QoS awareness; finally, the fourth group studies dynamic interference coordination without delay guarantees. As clearly evident, the categories show that there is a research gap on RRM decision making jointly considering \textit{heterogeneous QoS, including queue-aware delay sensitive flows, with dynamic interference coordination, and dynamic user association}.

\textbf{1 -} \emph{QoS without interference awareness}: This set of works studies the sub-problem of queue scheduling and resource allocation addressing only QoS and fairness without adequate attention to interference. Examples include maximizing average utilities balancing efficiency and fairness~\cite{SongTWC1}, analysis of generalized proportional fairness~\cite{Mo2000}, scheduling for elastic traffic using convex optimization~\cite{MadanTN2010Fast}, single cell throughput maximization with rate guarantees~\cite{Tao2008}, utility maximization with rate constraints through a token counter~\cite{AndrewsMinRate2005}, minimum rate guarantees using a Lagrangian approach~\cite{XWANG_IT}, joint channel- and queue-aware scheduling for mean-delay utility maximization~\cite{Song2009-CATA}, mean-delay fairness via gradient method~\cite{Alireza2011delay-fairness}, joint real-time and non-real-time packet scheduling and resource allocation~\cite{AlirezaTVT2014}, mean-delay guarantees through time-coupling constraints and Lagrange dual-based solutions~\cite{HajiesmailiArxive2015AverageDelayMultiHop}, maximizing goodput for multihop networks through dual solutions~\cite{SinghArxive2015AverageDelayMultiHop}, utility maximization and routing with probabilistic delay requirements~\cite{SaadWeiYuTNS2007DelayQoS}, adapting rates with delay constraints to increase network video capacity~\cite{GhoreishiAghvamiGC2015}, QoS-aware routing and subchannel allocation in time-slotted realy networks, without interference coordination, and dynamic user association~\cite{Hajipour2016}, and, in single AP, single frequency networks, optimizing secondary users' delay based on interference~\cite{EwaishaArxive2015AverageDelayForSU}.

\textbf{2 -} \emph{Static interference coordination with or without QoS awareness}: This set of works addresses a-priori static interference coordination, such as soft frequency reuse (SFR)~\cite{3GPPSoftFFR, MaoMaarefGC2008SFR}, two-phase coarse interference management and fine-scale resource-allocation based on graph-theoretic approaches~\cite{ChangTVT2009GraphForICIC}, joint optimization of user association and use of almost blank subframes (ABS)~\cite{MirshaCommLett2014JointABS}, single frame constant rate guarantees with interference threshold and a-priori user association~\cite{Zhang2014}, and outage guarantees on constant rate requirements with RRM in cognitive small cells using cooperative Nash bargaining~\cite{Zhang2015}. We note that~\cite{Tao2008, Zhang2014, Zhang2015, Zhang2016} consider delay constraints as constant rate guarantees; this approach is unsuitable for random arrivals with finite backlog (particularly relevant for delay sensitive flows).

\textbf{3 -} \emph{Dynamic interference coordination without QoS awareness}: This set of works aims at dynamic interference coordination for BE flows. Examples include throughput and fairness oriented interference coordination by blanking and based on dominant interferers~\cite{AkramTC2015ICIC}, dynamic interference avoidance for cell edge users~\cite{RahmanTWC2010ICIC}, a range of weighted sum signal-to-interference-plus-noise ratio (SINR) maximizations through Perron-Frobenius theory~\cite{LiangTWC2014NovelNUM}, and uplink clustering scheme decreasing both the intra- and inter-cluster interference without increasing the size of clusters \cite{HAJISAMI201744}. These interference coordination schemes improve cell-edge rates, but, importantly, do not address QoS.

\textbf{4 -} \emph{Dynamic interference coordination without delay guarantees}: This set focuses on rate QoS and interference coordination, without delay guarantees. Examples include energy efficiency maximization while guaranteeing minimum rates~\cite{LukaiGC2015ConcaveCovex, CoskunGC2015MinRateInterference}, interference management accounting for minimum throughputs with heterogeneous APs~\cite{AhujaSchaar2015ICIC}, hierarchic interference coordination with rate constraints~\cite{SadrTWC2014SmallCells}, load balancing and interference coordination with infinite backlog~\cite{WildmanICC2015PoorDelay} (note that infinite backlog assumption prevents the control of queue lengths), hybrid coordinated multipoint transmission, based on Markov decision process (MDP), improving the overall delay performances, but without delay guarantees~\cite{Li2017}, and heuristics for related sub-problems of interference coordination and queue equalization, based on static delays~\cite{MadanJSAC2010CellAssoc}. Among the studies considering interference and QoS, while rate metrics are useful with infinite backlog, they do not account for metrics such as delay, especially relevant for real-time flows.

\vspace{-\BeforeSectionTitle}
\subsection{Our Approach and Contributions}
\label{sec:NPhardGuess}
\vspace{-\AfterSectionTitle}
Our goal in this paper is to develop an algorithm enabling resources to follow users' traffic in a QoS- and interference-aware manner. Specifically, our goal is to associate users with APs and allocate time/frequency resources to maximize network utility while meeting rate and, crucially, \textit{delay constraints in a queue-aware manner, with random arrival}. Importantly, unlike the works reviewed in Section~\ref{sec:RelatedWorks}, we require joint dynamic interference coordination, dynamic user association, and finite backlog random arrivals (particularly relevant for delay sensitive flows).

This requirement implies that our algorithm(s) must meet five criteria: (1) We design for finite backlogs (queue-aware), because without queue-awareness, the algorithm cannot adapt frequency reuse to traffic situations. (2) We design for QoS- and load-awareness, accounting for (and exploiting) the heterogeneity of QoS classes, QoS requirements, and load conditions. (3) We design for network-wide interference awareness caused by frequency reuse. (4) We design for an opportunistic setup in order to exploit user, time, and frequency diversities. (5) Finally, if faced with an infeasible load condition, due to high mean input rates, or a spike in input rates, our design should allow for graceful degradation of the QoS satisfaction.

Meeting all these criteria is a complex problem: The problem is coupled across flows and across APs. Moreover, it is a nonlinear combinatorial program with a non-convex relaxed version. Unlike other works resulting in convex problems after integer relaxation (such as~\cite{Tao2008, Zhang2014, Zhang2015}), in which a time-sharing approach can be used, our problem, even for a single AP, does not lead to a convex problem, rather a convex \textit{maximization}, similar to \cite{Akram2013}. This is a strong indicator of an NP-hard problem~\cite{Raymond2010CovexMax}. We reemphasize that modeling of joint dynamic user association, dynamic interference, and finite backlog (unlike other works) inevitably lead to this challenging problem. With the global optimum essentially impossible to find, our \textsc{QoSaIC} algorithm allows us to meet the first above-mentioned four criteria while our \textsc{ILM} algorithm meets the fifth.

Having clarified the research gap and our approach, the contributions of this paper are:
\label{sec:Contributions}
\begin{enumerate}
\item We formulate a systematic multi-cell utility maximization problem, with heterogeneous QoS guarantees, including importantly \textit{queue-aware (finite backlog with random arrival)} delay sensitive flows. This enables matching (and relocating) of available time and frequency resources based on both the spatial dimension (user locations) and the temporal dimension (traffic arrivals). Unlike the related works (which use a constant rate constraint to serve delay sensitive flows), we consider the more realistic \textit{finite backlog with random arrivals}. Furthermore, again unlike related works, we do not use a static unilateral interference threshold or static user association, in order to not limit the network efficiency.

\item Unlike other works, our design allows for control of both instantaneous and mean QoS metrics. We derive a \textit{set of closed loop controls} that observe the mean QoS (rates and delays) measurements, compare them with the given QoS requirements, and adapt the allocation, in each frame. These controls are derived based on translating the high-level requirements to frame level requirements through Function \textsc{QoSiFT}, in Section~\ref{sec:Transformation}.

\item Given the channel and load information, our RRM makes decision jointly on time/frequency scheduling (QoS provisioning), short-term user association (including load balancing), and frequency reuse patterns (interference coordination), per frame. As extensively discussed in Section~\ref{sec:RelatedWorks}, our work is the first to address the above-mentioned RRM decisions jointly, to the best of our knowledge. This is done based on our several original techniques we develop in this paper~\footnote{namely, dealing with intra and inter cell interferences, shrinking the error tolerance on the inner loop, managing the error tolerance on the outer loop, novel primal and dual updates, and finally approximating the frugality constraint with Sigmoid function.}. Without these subtle techniques, the solution was not possible.

\item Unlike the related works, we also develop an effective strategy for \textit{infeasible load management} to account for scenarios with infeasible demands. This part of our solution enables us to have an RRM robust to the input load (not becoming infeasible due to input load) with graceful degradation.
\end{enumerate}

The paper is organized as follows: Section~\ref{sec:SysModelProblemFormulation} presents our system model, our novel formulation, and our novel translation of the high-level RRM to the frame level optimizations. Section~\ref{sec:Solution}, then, discusses the solution, based on several original techniques, developed in this paper, setting up the \textsc{QoSaIC} and \textsc{ILM} algorithms in Section~\ref{sec:algorithms}. Section~\ref{sec:simulations} presents the simulations to illustrate the effectiveness of our algorithms. Finally, Section~\ref{sec:conc} make the conclusions.

We use the conventional notation system: Boldface lower-case letters, e.g., $\mathbf{x}$, represent vectors, while boldface upper-case letters, e.g., $\mathbf{X}$, represent matrices. Calligraphy style letters are exclusively used for sets, e.g., $\mathcal{C}^{\cdot}$ with superscripts as required. Subscripts usually represent flow (user) and AP indices, while superscripts represent the frequency subchannel index, such as in $x_{\phi, p}^{(j)} [k]$~\footnote{Throughout the paper, we use this single format to index flow $\phi$, port $p$, and frequency $j$, such as in $x_{\phi,p}^{(j)} [k]$.}. We use $(j)$ in the superscript, so as to not confuse it with an exponent. When subscripts or superscript are used as \textit{mnemonic} to describe the nature of the quantity, we distinguish it using Roman style. Based on the convention, we distinguish functions by Fraktur letters, e.g., $\mathfrak{U} (\cdot)$. The mean value of a quantity is denoted by a bar, e.g., $\bar{r} [k]$. Finally, $[k]$ is exclusively used to denote the discrete-valued frame index $k$, similar to the convention of time series. Since the problem in this paper is inherently complex, it involves many notation. Thus, in Table~\ref{tab:symbols}, we summarize the symbols used throughout the paper, with their short definitions.
\begin{table}
\begin{center}
\small
\caption{List of symbols.} \label{tab:symbols}
\begin{tabular}{| p{.17\textwidth} | p{.27\textwidth} |}
\hline
\textbf{Symbols} & \textbf{Descriptions} \\ \hline \hline

\vspace{-0.05cm}
$J, P, \Phi$
& Dimensions of the problem, corresponding to indices $j, p, \phi$.
\\ \hline

\vspace{-0.05cm}
$T_{\mathrm{b}}, W_{\mathrm{b}}$
& Size of an RB, in sec., and in Hertz.
\\ \hline

\vspace{-0.05cm}
$\alpha, \beta, \nu$
& Algorithm fixed constants.
\\ \hline

\vspace{-0.05cm}
$x_{\phi, p}^{(j)} [k], \mathbf{X} [k]$
& Optimization main variable and its compact matrix representation.
\\ \hline

\vspace{-0.05cm}
$\mathsf{SINR}_{\phi, p}^{(j)} [k]$
& SINR on the link from user $\phi$, to port $p$, on RB $j$, in frame $k$.
\\ \hline

\vspace{-0.05cm}
$s_{\phi} [k], u_{\phi}^{(j)} [k], v_{p}^{(j)} [k], $
$ \mathbf{D} [k]$
& Dual variables and the matrix representation of them.
\\ \hline

\vspace{-0.05cm}
$\mathcal{F}^{\mathrm{DS}}, \mathcal{F}^{\mathrm{RS}}, \mathcal{F}^{\mathrm{BE}}$
& Set of flows in QoS classes: DS class, RS class, BE class.
\\ \hline

\vspace{-0.05cm}
$\mathcal{C}^{\mathrm{INT}}, \mathcal{C}^{\mathrm{{PHY}_1}}, \mathcal{C}^{\mathrm{{PHY}_2}}$, $\mathcal{C}^{\mathrm{{MAC}_{RS}}}, \mathcal{C}^{\mathrm{{MAC}_{DS}}}$, $\mathcal{C}^{\mathrm{f-MAC}}$
&Sets representing integer constraint, PHY-1 constraint, PHY-2 constraint, MAC constraint for RS class, MAC constraint for DS class, and MAC constraint translated into frame rates.
\\ \hline

\vspace{-0.05cm}
$\bar{d}_{\phi}^{\mathrm{max}}, \bar{r}_{\phi}^{\mathrm{min}}, \bar{r}_{\phi}^{\mathrm{max}}$
& Target QoS demands controlling $\bar{d}_{\phi} [k], \bar{r}_{\phi} [k]$.
\\ \hline

\vspace{-0.05cm}
$w_{\phi} [k]$, $r_{\phi} [k]$, $q_{\phi} [k]$, $\bar{q}_{\phi} [k]$, $a_{\phi} [k]$
& Fairness weight, frame rate, queue length, mean queue length, and instant arrivals.
\\ \hline

\vspace{-0.05cm}
$r_{\phi}^{\mathrm{{min}}}, r_{\phi}^{\mathrm{{max}}}$
& Translated minimum and maximum frame rates.
\\ \hline

\vspace{-0.05cm}
$\zeta^{(\cdot)}_{\phi} [k], \hbar [k]$
& Intermediate variables translating the mean-delay and mean-rate constraints to frame level optimizations.
\\ \hline

\vspace{-0.05cm}
$\gamma_{\phi, p}^{(j)} [k], \gamma_{\mathrm{noise}}$
&Channel coefficient, in frame $k$, noise level.
\\ \hline

\vspace{-0.05cm}
$\mathfrak{U}_{\phi} (\cdot), \mathfrak{Z}_{\phi} (\cdot), \mathfrak{T} (\cdot), $
$\mathfrak{L} (\cdot), \mathfrak{P}_{\phi, p}^{(j)}, \mathfrak{D} (\cdot), $
 $\mathfrak{D}^{\mathrm{s}}(\cdot), \mathfrak{D}^{\mathrm{u}}(\cdot), \mathfrak{D}^{\mathrm{v}}(\cdot)$, $\mathfrak{f} (\cdot)$
& Individual flow utility, Sigmoid function, overall translation, Lagrangian, overall primal update, overall dual update, dual update for variable $s$, dual update for variable $u$, dual update for variable $v$, AMC function.
\\ \hline

\vspace{-0.05cm}
$\mathrm{I}_{\phi, p}^{(j)} (\cdot)$
& Overall interference on link from user $\phi$ to port $p$ on RB $j$, in frame $k$, comprising of intercell interference $\mathrm{{I}^{\mathrm{inter}}}_{\phi, p}^{(j)} (\cdot)$ and intracell interference $\mathrm{{I}^{\mathrm{intra}}}_{\phi, p}^{(j)} (\cdot)$.
\\ \hline

\vspace{-0.05cm}
$\dagger_{p}^{(j)}, \ddagger_{p}^{(j)}, \Upsilon_{\phi, p}^{(j)}$
& Intermediate variables in the fixed point method algorithm.
\\ \hline

\vspace{-0.05cm}
$ \epsilon^{\mathrm{inner}}, \epsilon^{\mathrm{outer}}, g[k, i^{\mathrm{outer}}]$
& Error tolerances for inner/outer loop and the primal dual gap.
\\ \hline

\vspace{-0.05cm}
$\epsilon^{\mathrm{inner}}_{1}, \epsilon^{\mathrm{inner}}_{\infty}$
& Constants controlling the shrinkage of the inner loop error tolerance.
\\ \hline

\vspace{-0.05cm}
$\Delta^{s_{\phi}}, \Delta^{u_{p}^{(j)}}, \Delta^{v_{\phi}^{(j)}}$
&Satisfaction/violation margins for different constraints corresponding to the superscripted dual variables.
\\ \hline

\vspace{-0.05cm}
$\varpi, \aleph, \lambda^{\mathrm{max}}, \delta^{\mathrm{max}}$
& Constants associated with the novel dual update design.
\\ \hline

\vspace{-0.05cm}
$O^{\mathrm{r^{min}}}_{\phi} [k]$
& Single frame outage for rates.
\\ \hline

\vspace{-0.05cm}
$O^{\mathrm{\bar{r}^{min}}}_{\phi} [k], O^{\mathrm{\bar{d}^{max}}}_{\phi} [k],$ $\bar{O}^{\mathrm{\bar{d}^{max}}}_{\phi} [k]$
& Outage from mean QoS requirements corresponding to the superscripted QoS item (all for flow $\phi$, in frame $k$).
\\ \hline

\vspace{-0.05cm}
$i^{\mathrm{outer}}, i^{\mathrm{outer, max}},$ $i^{\mathrm{inner}}, i^{\mathrm{inner, max}}$
& Outer and inner loops counters and their corresponding allowed maximums.
\\ \hline
\end{tabular}
\end{center}
\vspace{-0.0cm}
\end{table}

\vspace{-\BeforeSectionTitle}
\section{System Model and High-Level Problem Formulation}
\label{sec:SysModelProblemFormulation}
\vspace{-\AfterSectionTitle}

\subsection{System Model}
We consider the downlink of a multi-cell OFDMA network comprising $P$ APs, serving $\Phi$ flows (users), \textit{without} a-priori user association. The available bandwidth is divided into $J$ resource blocks (RBs), each spanning $T_{\mathrm{b}}$ seconds and $W_{\mathrm{b}}$ Hertz. The system serves three classes of flows: a BE class, denoted by $\mathcal{F}^{\mathrm{BE}}$, comprising flows without rate or delay requirements, a DS class, $\mathcal{F}^{\mathrm{DS}}$, with a maximum mean-delay constraint for each flow $\phi$ ($\bar{d}_{\phi}^{\mathrm{max}}$), and a RS class, $\mathcal{F}^{\mathrm{RS}}$, with minimum mean-rate constraint for each flow ($\bar{r}_{\phi}^{\mathrm{min}}$); and without loss of generality, a maximum mean-rate constraint ($\bar{r}_{\phi}^{\mathrm{max}}$). The $P$ APs are connected to a C-RAN. The server also knows the data of all users and the channel state between all APs and all users, similar to other works in the fields, such as~\cite{Tao2008, Zhang2014, Zhang2015, Zhang2016}. In this paper, we provide solution for the air access. Backhaul scheduling remains as a future item extending this work.
\label{sec:AddedToChannelEstimation}

At the server, in frame $k$, each flow is associated with a queue of length $q_{\phi} [k]$ bits. The number of bits for flow $\phi$, that arrive in frame $k$, is denoted by $a_{\phi} [k]$. The product of the transmission power (with uniform transmit power allocation), antenna gain, and channel power, from AP $p$ to user $\phi$, on subchannel $j$, in frame $k$, is denoted by $\gamma^{(j)}_{\phi, p} [k]$, and is assumed known.

\vspace{-\BeforeSectionTitle}
\subsection{Components of High-Level Problem Formulation}
\label{sec:MainOptimization}
\vspace{-\AfterSectionTitle}
Our network objective is to maximize sum flow utilities, subject to the QoS constraints. The overall optimization problem is given in~\eqref{eq:optimVer1} on page~\pageref{eq:optimVer1}; we first develop optimization components of~\eqref{eq:optimVer1}, namely, optimization objective, optimization variable, interference metrics, flow rates, queuing delays, and optimization constraints.

\begin{itemize}
\item The optimization objective \eqref{eq:ObjectiveVer1} is the \emph{network utility}, $\mathfrak{U}_{\mathrm{T}} (\bar{\mathbf{r}} [k]) = \sum_{\phi=1}^{\Phi} \mathfrak{U}_{\phi} (\bar{r}_{\phi} [k])$, where $\bar{\mathbf{r}} [k]$ is the mean-rates vector. The objective is sum of the individual flow utilities, $\mathfrak{U}_{\phi} (\bar{r}_{\phi} [k])$, a function of individual mean-rate, $\bar{r}_{\phi} [k]$.

\item The main optimization variable is the binary $x^{(j)}_{\phi,p} [k]$: $x^{(j)}_{\phi,p} [k] =1$, if flow $\phi$ is scheduled to be served by AP $p$, on subchannel $j$, in frame $k$, else $x^{(j)}_{\phi,p} [k] =0$. Therefore, for each frame, the optimization variables form a 3D array denoted by $\mathbf{X}[k]$.

\item We now define the interference metrics in order to first calculate the signal-to-interference-plus-noise ratios (SINRs), and then find the rates on RBs. We denote the total interference impacting the link from AP $p$, supporting flow $\phi$, on subchannel $j$, by $\mathrm{I}_{\phi, p}^{(j)} (\mathbf{X} [k])$ comprising inter-cell and intra-cell interference.

The inter-cell interference to flow $\phi$, associated with AP $p$, on subchannel $j$, is due to undesired APs ($p' \neq p$) communicating on the same sub-channel:
\vspace{-\EqVspace}
\begin{multline}
\hspace{-0.7cm}
{\mathrm{I}^\mathrm{inter}}_{\phi, p}^{(j)}
(\mathbf{X} [k])
\triangleq
\sum_{{p'\neq p}}^{P}
\gamma^{(j)}_{\phi,p'} [k]
\sum_{\phi'=1}^{\Phi}
x_{\phi',p'}^{(j)} [k]
\\
=
\sum_{p'=1}^{P}
\gamma_{\phi, p'}^{(j)} [k]
\sum_{\phi'=1}^{\Phi}
x_{\phi', p'}^{(j)} [k]
-
\gamma_{\phi, p}^{(j)} [k]
\sum_{\phi'=1}^{\Phi}
x_{\phi',p}^{(j)} [k].
\label{eq:InterCellInterference}
\end{multline}

The intra-cell interference, on the same link, is given by
\begin{multline}
{\mathrm{I}^\mathrm{intra}}_{\phi, p}^{(j)}
(\mathbf{X} [k])
\triangleq
\gamma_{\phi, p}^{(j)} [k]
\sum_{{\phi' \neq \phi }}^{\Phi}
x_{\phi', p}^{(j)} [k] \\
=
\gamma_{\phi, p} ^{(j)} [k]
\sum_{\phi' =1}^{\Phi}
x_{\phi', p}^{(j)} [k]
-
\gamma_{\phi, p} ^{(j)} [k]
x_{\phi, p}^{(j)} [k].
\label{eq:IntraCellInterference}
\end{multline}
Intra-cell interference occurs when an AP serves more than one flow on a single RB. Later, we eliminate this totally undesirable situation via an explicit constraint~\footnote{Importantly, we note that without QoS constraints, accounting for intra-cell interference in the interference metric automatically eliminates it, i.e., a solution which allows for intra-cell interference cannot be a local optimum~\cite{ZheAdve}. With QoS constraints, on the other hand, this simplification is not valid. Furthermore, we note that this simplification in~\cite{ZheAdve} heavily depends on the assumption that the algorithm always \emph{guarantees} local optimality. This is not the case in many practical situations, when the algorithm cannot avoid trading off complexity with optimality. Therefore, we enforce the elimination of intra-cell interference through the explicit constraint of~\eqref{eq:constRBinCellPortsVer1}. We further elaborate on it when explaining the optimization constraints.}.
\label{sec:IntercellInterferenceExplanation}

Summing up the intra-cell and inter-cell interference metrics, the total interference is given by ${\mathrm{I}}_{\phi, p}^{(j)} (\mathbf{X} [k]) =
\Sigma_{p'=1}^{P}
\gamma_{\phi, p'}^{(j)} [k]
\Sigma_{\phi'=1}^{\Phi}
x_{\phi', p'}^{(j)} [k]
- \gamma_{\phi, p} ^{(j)} [k]
x_{\phi, p}^{(j)} [k]. \\
$

\item Based on the interference, the SINR, and the corresponding achievable spectral efficiency, are given by
\vspace{-\EqVspace}
\begin{align}
\vspace{-\EqVspace}
&\mathsf{SINR}_{\phi,p}^{(j)} [k]
\triangleq
\frac{{
x^{(j)}_{\phi,p} [k]
\gamma^{(j)}_{\phi,p} [k]
}}{
{\gamma_{\mathrm{noise}}
+
\mathrm{I}_{\phi, p}^{(j)} (\mathbf{X} [k]) }},
\;\; \\
&b_{\phi, p}^{(j)} [k] = \mathfrak{f}(\mathsf{SINR}_{\phi,p}^{(j)} [k]) = \log_2\left(1+\mathsf{SINR}_{\phi,p}^{(j)} [k]\right)
\label{eq:SINRandRateDefn}
\end{align}
explaining \eqref{eq:constInterferenceVer1}. Notations, $\gamma_{\mathrm{noise}}$ denotes the noise power and $\mathfrak{f}(\cdot)$ the capacity of the corresponding RB. Since any differentiable $\mathfrak{f}(\cdot)$ is allowed, an SINR gap to capacity can be added to~\eqref{eq:SINRandRateDefn}.
Having calculated the rate on RBs, the rate of a flow is given by~\eqref{eq:constFramebitrateVer1} summing up the RBs it is assigned.
We emphasize that, unlike e.g., \cite{Tao2008, Zhang2014, Zhang2015, Zhang2016}, we do not use a static interference threshold, but dynamic interference coordination. Furthermore, we do not use a-priori user association as in \cite{Tao2008, Zhang2014, Zhang2015}; our user association is part of $x_{\phi, p}^{(j)} [k]$ and changes from frame to frame.

\item We now describe the required constraints. The first constraint is on RB scheduling - \eqref{eq:constINTVer1} below. Furthermore, the physical layer imposes two constraints on any subchannel: first, frequency reuse is not allowed inside a cell - \eqref{eq:constRBinCellPortsVer1}; and second, a single flow cannot pass through two APs simultaneously over a single RB - \eqref{eq:constRBinCellFlowsVer1}. Note that while a flow cannot be connected to more than one AP, on a single RB, it \textit{can} be connected to multiple APs, across different RBs, allowing for \textit{data aggregation and load balancing}. As such, we emphasize that our formulation importantly allows for frequency reuse across APs. Since we use the joint approach, the frequency reuse adapts to channels and QoS requirements.

\item The QoS requirements, in~\eqref{eq:constMeanRateMaxMinVer1} and~\eqref{eq:constMeanDelayMaxVer1} below, represent the MAC constraints imposed as \emph{explicit} mean-rate and mean-delay constraints. BE flows do not impose QoS constraints. RS flows impose the constraints in~\eqref{eq:constMeanRateMaxMinVer1} while DS flows impose the constraints in~\eqref{eq:constMeanDelayMaxVer1}, where $\bar{d}_\phi[k]$ and $\bar{r}_\phi[k]$ denote the \emph{mean}-delay and \emph{mean}-rate achieved by flow $\phi$, respectively. The relation in~\eqref{eq:constMeanbitrateVer1} relates the mean-rate to the \emph{instantaneous} rate, in frame $k$, $r_\phi[k]$, using forgetting factor $\hbar [k]$. We discuss the connection of mean and frame quantities, in Section~\ref{sec:Transformation}.
\end{itemize}

Having explained the optimization objective, variables, and the constraints, our core proposed optimization problem is given in~\eqref{eq:optimVer1}.

\vspace{-0.5cm}
\begin{subequations}
\label{eq:optimVer1}
\begin{align}
&\underset{x_{\phi,p}^{(j)} [k]}{\text{max}}
\quad
\mathfrak{U}_{\mathrm{T}} (\bar{\mathbf{r}} [k])
\triangleq
\sum_{\phi=1}^{\Phi}
\mathfrak{U}_{\phi}
(\bar{r}_{\phi} [k])
\label{eq:ObjectiveVer1}
\end{align}
\vspace{-0.5cm}
\begin{align}
&
\mathcal{C}^{\mathrm{INT}}
\triangleq
\{
\forall \phi, p, j: x_{\phi, p}^{(j)} [k] \in \{0, 1\}
\},
\label{eq:constINTVer1}\\
&
\mathcal{C}^{\mathrm{PHY_1}}
\triangleq
\{
\forall j,p:
{\Sigma_{\phi=1}^{\Phi} x_{\phi,p}^{(j)} [k] \leq 1}
\},
\label{eq:constRBinCellPortsVer1}\\
&
\mathcal{C}^{\mathrm{PHY_2}}
\triangleq
\{
\forall j,\phi:
{\Sigma_{p=1}^{P} x_{\phi,p}^{(j)} [k] \leq 1}
\},
\label{eq:constRBinCellFlowsVer1}\\
&
\mathcal{C}^{\mathrm{MAC_{RS}}} [k]
\triangleq
\{
\forall \phi:
\bar{r}_{\phi}^{\mathrm{min}}
\leq
\bar{r}_{\phi} [k]
\leq
\bar{r}_{\phi}^{\mathrm{max}}
\},
\label{eq:constMeanRateMaxMinVer1} \\
&
\mathcal{C}^{\mathrm{MAC_{DS}}} [k]
\triangleq
\{
\forall \phi:
\bar{d}_{\phi} [k]
\leq
\bar{d}_{\phi}^{\mathrm{max}}
\},
\label{eq:constMeanDelayMaxVer1} \\
& \forall \phi:
\bar{r}_{\phi} [k]
\triangleq
(1-\hbar [k])
\bar{r}_{\phi} [k-1]
+
\hbar [k] r_{\phi} [k],
\label{eq:constMeanbitrateVer1}\\
&
\forall \phi:
r_{\phi} [k]
\triangleq
W_{\mathrm{b}}
\Sigma_{p=1}^{P}
\Sigma_{j=1}^{J}
b_{\phi,p}^{(j)} [k],
\label{eq:constFramebitrateVer1}\\
&
\forall \phi,j,p:
b_{\phi,p}^{(j)} [k]
=
\mathfrak{f}
\big(\mathsf{SINR}_{\phi,p}^{(j)} [k]\big).
\label{eq:constInterferenceVer1}
\end{align}
\end{subequations}

To the best of our knowledge (and as extensively reviewed in Section~\ref{sec:RelatedWorks}), the formulation in~\eqref{eq:optimVer1} is the first one incorporating \textit{delay and rate QoS} as explicit constraints, on a \textit{multi-frame} problem, while also accounting for \textit{dynamic interference} and \textit{dynamic user association}. Unlike other works, we consider \textit{finite backlog} resulting in queue awareness and addressing \textit{time varying random arrivals}, crucial for delay sensitive flows. This formulation makes our first major contribution, summarized in Section~\ref{sec:Contributions}. In the next part, we translate the problem in~\eqref{eq:optimVer1} into a parameterized frame-by-frame rate optimization problems. We highlight that the optimization in~\eqref{eq:optimVer1} is executed for every frame $k$.

\vspace{-\BeforeSectionTitle}
\section{Translation to Frame-level}
\label{sec:Transformation}
\vspace{-\AfterSectionTitle}
\subsection{Translating MAC Constraints to Frame Rate Constraints}
\vspace{-\AfterSectionTitle}
We begin with the RS flows. The instantaneous rate, in frame $k$, is given by~\eqref{eq:constFramebitrateVer1}, aggregating all the RBs given to a link. Mean-rate is calculated using exponential averaging (with averaging coefficients $\hbar [k]$) in~\eqref{eq:constMeanbitrateVer1}. We use $\hbar [k] = 1/k$. After simple manipulation of the inequalities (substituting~\eqref{eq:constFramebitrateVer1} into~\eqref{eq:constMeanbitrateVer1} and solving for frame rate $r_{\phi} [k]$), the constraints on the minimum and maximum mean-rates translate to single \emph{frame} constraints as in
\vspace{-\EqVspace}
\vspace{-\EqVspace}
\begin{align*}
\vspace{-\EqVspace}
\vspace{-\EqVspace}
\bar{r}_{\phi}^{\mathrm{min}}
&\leq
\bar{r}_{\phi} [k]
\Leftrightarrow
\\
&r^{\mathrm{min_{1}}}_{\phi} [k]
\triangleq
\big(
{\bar{r}_{\phi}^{\mathrm{min}} - (1-\hbar [k]) \bar{r}_{\phi} [k-1]}
\big)
/
{\hbar [k]}
\leq
r_{\phi} [k].
\\
\bar{r}_{\phi} [k]
&\leq
\bar{r}_{\phi}^{\mathrm{max}}
\quad
\Leftrightarrow
\\
\quad
&r_{\phi} [k]
\;
\leq
\;
r^{\mathrm{max_{1}}}_{\phi} [k]
\triangleq
\big(
{\bar{r}_{\phi}^{\mathrm{max}} - (1-\hbar [k]) \bar{r}_{\phi} [k-1]}
\big)
/
{\hbar [k]}.
\end{align*}

\label{sec:TranslatingDS}
We now translate the mean-delay requirements into frame rates requirements. In contrast to mean-rate, mean-delay requires a more detailed analysis. Using Little's formula ($\bar{d}_{\phi} [k] \approx \bar{q}_{\phi} [k] \big/ \bar{r}_{\phi} [k]$), queue evolution (conservation on arrivals and departures), and estimating the arrival through its empirical expected value, the mean-delay can be approximated as (see~\cite{Alireza2011delay-fairness}):
\vspace{-\EqVspace}
\begin{equation}
\vspace{-\EqVspace}
\bar{d}_{\phi} [k] \approx
\frac{\zeta_{\phi}^{(1)} [k] - \zeta_{\phi}^{(2)} [k] r_{\phi} [k]}
{\zeta_{\phi}^{(3)} [k] + \zeta_{\phi}^{(4)} [k] r_{\phi} [k]},
\label{eq:MeanDelayEst}
\end{equation}
where the approximation is valid when there is enough backlog in the queue:
\vspace{-\EqVspace}
\vspace{-\EqVspace}
\begin{equation}
\vspace{-\EqVspace}
\vspace{-\EqVspace}
r_{\phi} [k]
\;
\leq
\;
r^{\mathrm{max_{2}}}_{\phi} [k]
\triangleq
{(q_{\phi} [k-1]+ T_{\mathrm{b}} \bar{r}_{\phi} [k-1])}
/
{T_{\mathrm{b}}}.
\label{eq:FrugalityConstraint}
\end{equation}
This \emph{frugality constraint} ensures that the service rate to be less than the backlog and prevents resources being wasted on a flow without a sufficient backlog. As derived in~\cite{Alireza2011delay-fairness}, the constants $\zeta^{(\cdot)}_{\phi} [k]$ are functions of previous queue lengths and service rates:
\vspace{-\EqVspace}
\begin{equation*}
\vspace{-\EqVspace}
\begin{split}
&
\zeta^{(1)}_{\phi} [k]
=
\frac{(k-2)}{k} \bar{q}_{\phi} [k-2]
+
\frac{2}{k} q_{\phi} [k-1]
+
\frac{T_{\mathrm{b}} (k-1)}{k^2} \bar{r}_{\phi} [k-1], \\
&
\zeta^{(2)}_{\phi} [k]
=
T_{\mathrm{b}} {(k-1)}/{k^2},\\
&
\zeta^{(3)}_{\phi} [k]
=
{(k-1)} \bar{r}_{\phi} [k-1]/{k},
\; \;
\zeta^{(4)}_{\phi} [k]
=
{1}/{k},
\end{split}
\end{equation*}
where the mean queue length is denoted by $\bar{q}_{\phi} [k]\triangleq \Sigma_{k'=1}^{k} q_{\phi} [k'] / k$.
Since~\eqref{eq:MeanDelayEst} represents a decreasing function of $r_{\phi} [k]$, constraint on mean-delay in~\eqref{eq:constMeanDelayMaxVer1}, combined with~\eqref{eq:MeanDelayEst}, yields to
\vspace{-\EqVspace}
\vspace{-\EqVspace}
\begin{equation}
\vspace{-\EqVspace}
r_{\phi}^{\mathrm{min_{2}}} [k] \triangleq
\frac{\zeta_{\phi}^{(1)} [k] - \zeta_{\phi}^{(3)} [k] \bar{d}^{\mathrm{max}}_{\phi}}
{\zeta_{\phi}^{(2)} [k] + \zeta_{\phi}^{(4)} [k] \bar{d}^{\mathrm{max}}_{\phi}} \leq r_{\phi} [k],
\label{eq:Rmin2FromDelay}
\end{equation}
where $r_{\phi}^{\mathrm{min_{2}}} [k]$ is now a second min constraint on frame rates due to the delay constraint (the first minimum rate, $r_{\phi}^{\mathrm{min_{1}}} [k]$, was due to RS flows). We highlight that, unlike other works \cite{Tao2008, Zhang2014, Zhang2015, Zhang2016}, a constant minimum rate guarantees across frames is not sufficient for DS flows. Instead, as derived here, the intricate function in~\eqref{eq:Rmin2FromDelay} is needed.

We highlight that in this paper we choose to use the mean-delay for the DS flows, similar to~\cite{Song2004-CATA}. Using other metrics, particularly head-of-the-line (HOL) delay~\cite{AlirezaTVT2014} makes the problem highly complex in terms of connecting the optimization variable to the delay metric. Nevertheless, interestingly, since we guarantee a bound on mean-delay, we also implicitly \textit{guarantee a probabilistic bound on HOL}. Based on Markov inequality, to control the outage on HOL-delay bound, we can perform it through a bound on mean-delay outage. In other words, bounding mean-delay to $\bar{d}_{\phi}^{\mathrm{max}} = d^{\mathrm{HOL^{max}}}_{\phi} \delta_{\phi}$ bounds the HOL-outage at most to $\delta_{\phi}$.

Having translated the MAC constraints, we see that the \emph{mean-rate} and the \emph{mean-delay} constraints are equivalent to two independent min \emph{frame rate} constraints and two independent max \emph{frame rate} constraints. These four constraints simplify to a single minimum of $r^{\mathrm{min}}_{\phi} [k] \triangleq \max(r_{\phi}^{\mathrm{min_{1}}} [k],r_{\phi}^{\mathrm{min_{2}}} [k])$ and a single maximum of $ r^{\mathrm{max}}_{\phi} [k] \triangleq \min(r^{\mathrm{max_{1}}}_{\phi} [k],r^{\mathrm{max_{2}}}_{\phi} [k])$.
We denote the feasible set of these \textit{frame requirements} as $\mathcal{C}^{\mathrm{f-MAC}} [k]$, while we keep using $\mathcal{C}^{\mathrm{MAC}} [k]$ for the feasible set of \textit{mean requirements}.

\vspace{-\BeforeSectionTitle}
\subsection{Linearizing the Objective Function}
\label{sec:RefiningTheObjective}
\vspace{-\AfterSectionTitle}
We now focus on the objective in~\eqref{eq:optimVer1}. Following the common practice in resource allocation literature (e.g. see~\cite[9.3.2]{WeiYuBookForTaylor}), we use a Taylor expansion to linearize the objective with respect to the previous frame. Using~\eqref{eq:constMeanbitrateVer1}, the $1^\text{st}$ order Taylor series yields
\vspace{-\EqVspace}
\begin{multline}
\vspace{-\EqVspace}
\hspace{-0.2cm}
\max \sum_{\phi = 1}^{\Phi} \mathfrak{U}_{\phi} (\bar{r}_{\phi} [k])
\approx \\
\sum_{\phi = 1}^{\Phi}
\mathfrak{U}_{\phi}
\big(
(
1-\hbar [k]
)
\bar{r}_{\phi} [k-1]
\big)
+
\hbar [k]
\max
\sum_{\phi=1}^{\Phi}
w_{\phi}[k]
r_{\phi}[k].
\vspace{-\EqVspace}
\vspace{-0.5cm}
\end{multline}
The \emph{fairness weights} are given by
$w_{\phi}[k] \triangleq \frac{\partial \mathfrak{U}_{\phi} (R)} {\partial R} $, at $R = (1-\hbar [k]) \bar{r}_{\phi} [k-1]$. With the Taylor expansion, sum utility is approximated by maximization of a weighted sum of frame rates. We note that utilities are concave increasing functions in order to model \emph{diminishing marginal utility}.

\vspace{-\BeforeSectionTitle}
\subsection{Approximating the Frugality Constraint with a Soft Constraint}
\label{sec:SoftFrugalityConstraint}
\vspace{-\AfterSectionTitle}
Here, we particularly consider the feasibility of the frugality constraint.
In low load conditions, the frugality constraint~\eqref{eq:FrugalityConstraint} often makes the requirements on frame rates infeasible; this is especially a problem when we have a low backlog and/or when RB granularity does not match the needed frame rates. We therefore approximate the frugality constraint with a soft constraint inside the optimization objective. In~\cite{Song2004-CATA}, the authors substitute the weighted sum objective with $\Sigma_{\phi=1}^{\Phi} w_{\phi} \min (r_{\phi} [k], r^{\mathrm{\max}}_{\phi} [k])$. However, since we want to use a fixed point method for the primal problem, we must resolve the discontinuity in the derivative of $\min(r_{\phi} [k], r^{\mathrm{\max}}_{\phi} [k])$. Therefore, we approximate this soft constraint with a Sigmoid function as
\begin{equation}
\hspace{-0.125cm}
\min (r_{\phi} [k], r^{\mathrm{\max}}_{\phi} [k])
\approx
\mathfrak{Z}_{\phi} (r_{\phi})
\triangleq
\frac{1}{\nu}
\log
\left(
\frac{ \mathrm{e}^{\nu(r_{\phi} [k] - r_{\phi}^{\mathrm{max}} [k])}}
{1 + \mathrm{e}^{\nu(r_{\phi} [k] - r_{\phi}^{\mathrm{max}} [k])} }
\right),
\label{eq:SigmoidIntegral}
\end{equation}
where the parameter $\nu$ controls the sharpness of the Sigmoid function (we use $\nu= 0.1$).

Note that we cannot keep the frugality constraint as an explicit constraint, because there are backlog scenarios, where including the constraint explicit makes the problem infeasible. Having an infeasible problem, especially only due to the frugality constraint is not acceptable, in terms of robustness of the formulation to the input loads. We selected using Sigmoid approach after implementing the explicit frugality constraints, understanding its limitations, comparing alternative approaches for the soft constraints, and finally selecting the most effective one.

We emphasize that approximating the frugality constraint inside the objective \textit{does not} impact the heterogeneous QoS guarantees nor the dynamic interference coordination capabilities. Moreover, this constraint is active only in low load conditions, and is \textit{inactive} in moderate and high load conditions, anyways. Our optimization, even without frugality constraint is valid. The reason is that when the allocated data from a flow exceeds the actual backlog, one can always cut the surplus. However, we include the frugality constraint in order to open up space more efficiently for BE flows, and push BE flows further in service (whenever possible).
\label{sec:AddedToFrugality}

\vspace{-\BeforeSectionTitle}
\subsection{Summary of Translation}
\vspace{-\AfterSectionTitle}
The function QoS inter-frame translator (\textsc{QoSiFT}) described below summarizes the discussed translations so far. It represents a \textit{set of closed loop controls} adjusting the frame requirements, based on comparing the measurements with the targets, in order to maintain the RS and DS users' satisfaction while serving as many BE flows as possible.
\label{sec:QoSift}
\vspace{-\EqVspace}
\begin{algorithm}[H]
\floatname{algorithm}{Function}
\renewcommand{\thealgorithm}{}
\caption{\textsc{QoSiFT}
$\big(
{q}_{\phi} [k-1],
{r}_{\phi} [k-1],
\bar{r}_{\phi}^{\mathrm{min}},
\bar{r}_{\phi}^{\mathrm{max}},
\bar{d}_{\phi}^{\mathrm{max}}
\big)$
$.\quad \quad \quad\quad \quad \quad \quad \quad \rightarrow \quad \quad
\big(
r_{\phi}^{\mathrm{min}} [k],
r_{\phi}^{\mathrm{max}} [k],
w_{\phi} [k]
\big)
$
}
\begin{algorithmic}[1]
\State
$
r^{\mathrm{min}}_{\phi} [k]
\leftarrow
\max
\big(
\frac{\bar{r}_{\phi}^{\mathrm{min}} - (1-\hbar [k]) \bar{r}_{\phi} [k-1]}
{\hbar [k]}
,
\frac{\zeta_{\phi}^{(1)} [k] - \zeta_{\phi}^{(3)} [k] \bar{d}^{\mathrm{max}}_{\phi}}
{\zeta_{\phi}^{(2)} [k] + \zeta_{\phi}^{(4)} [k] \bar{d}^{\mathrm{max}}_{\phi}}
\big)
$.
\State
$
r^{\mathrm{max}}_{\phi} [k]
\leftarrow
\min
\big(
\frac{\bar{r}_{\phi}^{\mathrm{max}} - (1-\hbar [k]) \bar{r}_{\phi} [k-1]}
{\hbar [k]}
,
\frac{q_{\phi} [k-1]+ T_{\mathrm{b}} \bar{r}_{\phi} [k-1]}
{T_{\mathrm{b}}}
\big).
$
\State
$
w_{\phi}[k]
\leftarrow
\frac{\partial \mathfrak{U}_{\phi} (R)}
{\partial R}
\;
\text{at}
\;
{R = (1-\hbar [k]) \bar{r}_{\phi} [k-1]}.
$
\end{algorithmic}
\end{algorithm}

We now summarize the original and the translated formulations in~\eqref{eq:optimVerRepeatedOriginal} and~\eqref{eq:optimVerFinal}, respectively.
\begin{figure}[H]
\vspace{-\EqVspace}
\vspace{-\EqVspace}
\begin{minipage}{0.45\linewidth}
\begin{subequations}
\label{eq:optimVerRepeatedOriginal}
\begin{align}
&\underset{x_{\phi,p}^{(j)} [k]}{\text{max}}
\sum_{\phi=1}^{\Phi} \mathfrak{U}_{\phi} (\bar{r} [k]). \\
& { \mathbf{X} [k] \in \mathcal{C}^{\mathrm{PHY}}, }\\
& { \mathbf{r} [k] = \mathfrak{T} (\mathbf{X} [k]), }\\
& { \mathbf{\bar{r}} [k] \in \mathcal{C}^{\mathrm{{MAC}_{RS}}} [k], } \\
& { \mathbf{\bar{d}} [k] \in \mathcal{C}^{\mathrm{{MAC}_{DS}}} [k]. }
\end{align}
\end{subequations}
\end{minipage}
\hspace{0.2cm}
\begin{minipage}{0.45\linewidth}
\begin{subequations}
\label{eq:optimVerFinal}
\begin{align}
&\underset{x_{\phi,p}^{(j)} [k]}{\text{max}}
\sum_{\phi=1}^{\Phi} w_{\phi} \mathfrak{Z}_{\phi} (r_{\phi} [k]). \\
& { \; \mathbf{X} [k] \in \mathcal{C}^{\mathrm{PHY}}, }\\
& { \; \mathbf{r} [k] = \mathfrak{T} (\mathbf{X} [k]), }\\
& { \; \mathbf{r}^{\mathrm{min}} [k] \leq \mathbf{r} [k].}
\end{align}
\end{subequations}
\end{minipage}
\vspace{-\EqVspace}
\vspace{-\EqVspace}
\vspace{-\EqVspace}
\end{figure}

\noindent Here, $\mathfrak{T} (\cdot)$ represents~\eqref{eq:constFramebitrateVer1} and~\eqref{eq:constInterferenceVer1}.
Symbol $\mathcal{C}^{\mathrm{PHY}}$ denotes the intersection of PHY layer constraints: $\mathcal{C}^{\mathrm{PHY}} = \mathcal{C}^{\mathrm{PHY}_1} \cap \mathcal{C}^{\mathrm{PHY}_2} \cap \mathcal{C}^{\mathrm{INT}}$.
The other notations used here are as in~\eqref{eq:optimVer1}.
The translation in this section forms our second major contribution, discussed in Section~\ref{sec:Contributions}.
We note that, unlike many of the related works, our problem in~\eqref{eq:optimVerFinal} is not convex (due to dynamic interference coordination and dynamic user association). As such, in the next section, we combine several techniques in order to devise an effective solution.

\vspace{-\BeforeSectionTitle}
\section{Solution Approach}
\label{sec:Solution}
\vspace{-\AfterSectionTitle}
Having formulated the RRM problem, we now use a tailored primal-dual approach, coupled with a number subtle novel techniques, to devise an effective solution. Properly revised versions of the primal-dual method have proven to be powerful to devise approximation algorithms for combinatorial optimization, e.g., Hungarian algorithm~\cite{Vazirani2001, BertsekasCons}. We first relax the integer constraint and decompose the problem into primal and dual domains. We then use the Karush Kuhn Tucker (KKT) conditions to form a system of equations for the primal domain. Next, we solve the primal problem using fixed point iterations (in an \emph{inner} loop) while also using a novel approach to update the dual values (in an \emph{outer} loop). The integer constraint is imposed iteratively, with a \emph{primal dual interface} projection, for each outer iteration.

\vspace{-\BeforeSectionTitle}
\subsection{Decomposing into Primal-Dual Domains}
\vspace{-\AfterSectionTitle}
The Lagrangian of the relaxed constrained problem in~\eqref{eq:optimVerFinal} is given by
\vspace{-\EqVspace}
\begin{equation}
\vspace{-\EqVspace}
\normalsize
\begin{split}
\hspace{-0.2cm}
&\mathfrak{L}
\big(
\mathbf{X},
\mathbf{D}
\big)
 =
\sum_{\phi=1}^{\Phi}
\big(
w_{\phi} \mathfrak{Z}_{\phi} (r_{\phi})
-
s_{\phi} (r_{\phi}^{\mathrm{min}} - r_{\phi})
\big)
\\
&
- \sum_{j=1}^{J}
\sum_{p=1}^{P}
\sum_{\phi=1}^{\Phi}
x_{\phi,p}^{(j)}
(u_{p}^{(j)} + v_{\phi}^{(j)})
-
\frac{u_{p}^{(j)}}{\Phi}
-
\frac{v_{\phi}^{(j)}}{P},
\end{split}
\label{eq:Lagrangian}
\end{equation}
where
$s_{\phi},
u_{p}^{(j)}, v_{\phi}^{(j)}$
are the Lagrangian multipliers associated with the minimum rate requirement,
$\mathcal{C}^{\mathrm{PHY}_1}$, and $\mathcal{C}^{\mathrm{PHY}_2}$ constraints, respectively. Vector versions of the Lagrangian multipliers are denoted by
$\mathbf{s},
\mathbf{u},
\mathbf{v}$, and
$\mathbf{D}
\triangleq
(
\mathbf{s},
\mathbf{u},
\mathbf{v}
)$.
For simplicity, in the derivations, we drop the frame index $k$. The constrained optimization now becomes an unconstrained problem~\cite{BertsekasCons}, as in
\vspace{-\EqVspace}
\vspace{-\EqVspace}
\begin{equation}
\vspace{-\EqVspace}
\min_{ 0 \leq \mathbf{D} }
\sup_{ \mathbf{X} }
\mathfrak{L}
(
\mathbf{X},
\mathbf{D}
),
\label{eq:PrimalDualDecomp}
\end{equation}
where $0\leq\mathbf{D}$ is element-wise. Forming the dual problem suggests an iterative solution between the primal and dual domains: iterate between solving $\sup_{\mathbf{X}} \mathfrak{L} (\mathbf{X}, \mathbf{D})$ in order to find the primal variables and solving $\min_{0\leq\mathbf{D}} \sup_{\mathbf{X}} \mathfrak{L} (\mathbf{X}, \mathbf{D})$ in order to find the dual variables.

With explicit QoS constraints, an insightful interpretation is that the Lagrangian is equivalent to solving a multi-objective optimization, where, in addition to the conventional objective, the other objectives satisfy the QoS constraints. The Lagrangian jointly finds the appropriate scale factors to this multi-objective optimization. This is in contrast to including the constraints within the objective function, where the scale factors have to be adjusted manually. Works based on QoS constraints inside the objective have reported difficulty adjusting these scale factors~\cite{AndrewsMinRate2005}.

\vspace{-\BeforeSectionTitle}
\subsection{Dealing with Primal Variables}
\label{sec:PrimalDomain}
\vspace{-\AfterSectionTitle}
In this section, we solve $\sup_{\mathbf{X}} \mathfrak{L} (\mathbf{X}, \mathbf{D})$ for the primal variables, assuming fixed dual variables. Vanishing the derivative with respect to the primal variables yields to
\vspace{-\EqVspace}
\begin{equation}
\vspace{-\EqVspace}
\frac{\partial \mathfrak{L}}
{\partial x_{\hat{\phi},\hat{p}}^{(\hat{j})}}
=
\sum_{\phi=1}^{\Phi}
\bigg(
\frac{\partial r_{\phi}}{\partial x_{\hat{\phi},\hat{p}}^{(\hat{j})}}
\Big(
w_{\phi} \frac{\partial \mathfrak{Z}_{\phi}(r_{\phi})}{\partial r_{\phi}}
+ s_{\phi}
\Big)
\bigg)
-
u_{\hat{p}}^{(\hat{j})}
-
v^{(\hat{j})}_{\hat{\phi}} =0.
\label{eq:DerivativeToX}
\end{equation}
We denote the derivative of the Sigmoid function in~\eqref{eq:SigmoidIntegral} as $\mathfrak{s}_{\phi} (r_{\phi}) \triangleq {\partial \mathfrak{Z}_{\phi} (r_{\phi})} / {\partial r_{\phi}} $ leading to $\mathfrak{s}_{\phi}(r_{\phi}) = {1} \big/ ({1 + \mathrm{e}^{-\nu(r_{\phi}^{\mathrm{max}} [k] - r_{\phi} [k])} })$. The derivative of the Lagrangian has three components, $\Upsilon^{(\hat{j})}_{\hat{\phi}, \hat{p}}, \dagger^{(\hat{j})}_{\hat{p}}, \ddagger^{(\hat{j})}_{\hat{\phi}, \hat{p}}$, arising from the derivative of $r_{\phi}$. Considering the fact that the derivative of rate can be written as in \eqref{eq:OriginalDerivative}, the conditions in~\eqref{eq:DerivativeToX} dictates $\Upsilon^{(\hat{j})}_{\hat{\phi}, \hat{p}} + \dagger^{(\hat{j})}_{\hat{p}} + \ddagger^{(\hat{j})}_{\hat{\phi}, \hat{p}} - u_{\hat{p}}^{(\hat{j})} - v^{(\hat{j})}_{\hat{\phi}} = 0$, where the components, $\Upsilon^{(\hat{j})}_{\hat{\phi}, \hat{p}}, \dagger^{(\hat{j})}_{\hat{p}}, \ddagger^{(\hat{j})}_{\hat{\phi}, \hat{p}}$, are as below.

\begin{figure*}[t]
\normalsize
\setcounter{MYtempeqncnt}{\value{equation}}
\setcounter{equation}{15}
\begin{equation}
\hspace{-0.3cm}
\frac{\partial r_{\phi}}
{\partial x_{\hat{\phi},\hat{p}}^{(\hat{j})}}
=
W_{\mathrm{b}}
\sum_{j =1}^{J}
\sum_{p =1}^{P}
\Bigg(
\overbrace{
\frac{
\frac{ \partial
(
\gamma_{\phi,p}^{(j)}
x_{\phi,p}^{(j)}
)
}
{{\partial x_{\hat{\phi},\hat{p}}^{(\hat{j})}}}
}
{
\big(
\gamma_{\mathrm{n}}
+
\mathrm{I}_{\phi,p}^{(j)}
+
\gamma_{\phi,p}^{(j)} x_{\phi,p}^{(j)}
\big)
}
}^{\text{related to $\Upsilon^{(\hat{j})}_{\hat{\phi}, \hat{p}}$}}
-
\frac{
\overbrace{
\gamma_{\phi,p}^{(j)} x_{\phi,p}^{(j)}
\frac{\partial \mathrm{{I}^{inter}}_{\phi,p}^{(j)}}
{{\partial x_{\hat{\phi},\hat{p}}^{(\hat{j})}} }
}^{\text{related to $\dagger^{(\hat{j})}_{\hat{p}}$}}
+
\overbrace{
\gamma_{\phi,p}^{(j)} x_{\phi,p}^{(j)}
\frac{\partial \mathrm{{I}^{intra}}_{\phi,p}^{(j)}}
{{\partial x_{\hat{\phi},\hat{p}}^{(\hat{j})}} }
}^{\text{related to $\ddagger^{(\hat{j})}_{\hat{\phi}, \hat{p}}$}}
}
{
\big(
\gamma_{\mathrm{n}}
+
\mathrm{I}_{\phi,p}^{(j)}
+
\gamma_{\phi,p}^{(j)} x_{\phi,p}^{(j)}
\big)
\big(
\gamma_{\mathrm{n}}
+
\mathrm{I}_{\phi,p}^{(j)}
\big)
}
\Bigg),
\label{eq:OriginalDerivative}
\end{equation}
\vspace{-0.5cm}
\end{figure*}
The first component ($\Upsilon^{(\hat{j})}_{\hat{\phi}, \hat{p}}$) is a single term depending on $x_{\hat{\phi},\hat{p}}^{(\hat{j})}$ as
\vspace{-\EqVspace}
\begin{equation}
\vspace{-\EqVspace}
{\color{black}
\Upsilon^{(\hat{j})}_{\hat{\phi}, \hat{p}}
}
\triangleq
\big(
w_{\hat{\phi}} \mathfrak{s}_{\hat{\phi}}(r_{\hat{\phi}})
+ s_{\hat{\phi}}
\big)
W_{\mathrm{b}}
\frac{\gamma_{\hat{\phi}, \hat{p}}^{(\hat{j})}
}
{
\gamma_{\mathrm{n}}
+
\mathrm{I}_{\hat{\phi}, \hat{p}}^{(\hat{j})}
+
\gamma_{\hat{\phi}, \hat{p}}^{(\hat{j})} x_{\hat{\phi},\hat{p}}^{(\hat{j})}
}.
\label{eq:Upsilon}
\end{equation}
The second component ($\dagger^{(\hat{j})}_{\hat{p}}$) is due to inter-cell interference. The derivative of $\mathrm{{I}^{inter}}_{\phi,p}^{(j)}$ has non-zero elements only when $p \neq \hat{p}, j=\hat{j}$, and is equal to $\gamma^{(\hat{j})}_{{\phi},\hat{p}}$.
This yields $\Phi(P-1)$ terms:
\vspace{-\EqVspace}
\begin{equation}
\vspace{-\EqVspace}
\dagger^{(\hat{j})}_{\hat{p}}
(\mathbf{X}, \mathbf{D})
\triangleq
\sum_{\phi=1}^{\Phi}
\sum_{\substack{ p \neq \hat{p}}}^{P}
W_{\mathrm{b}}
\frac{
- (
w_{{\phi}} \mathfrak{s}_{\phi}( r_{\phi})
+ s_{{\phi}}
)
\gamma^{(\hat{j})}_{{\phi},{p}}
x^{(\hat{j})}_{{\phi},{p}}
\gamma^{(\hat{j})}_{{\phi},\hat{p}}
}
{
\big(
\gamma_{\mathrm{n}}
+
\mathrm{I}_{{\phi},p}^{(\hat{j})}
+
\gamma_{{\phi},p}^{(\hat{j})}
x_{{\phi},p}^{(\hat{j})}
\big)
\big(
\gamma_{\mathrm{n}}
+
\mathrm{I}_{{\phi},p}^{(\hat{j})}
\big)
}.
\label{eq:DaggerOne}
\end{equation}
The third component ($\ddagger^{(\hat{j})}_{\hat{\phi}, \hat{p}}$) is based on intra-cell interference. The derivative of $\mathrm{{I}^{intra}}_{\phi,p}^{(j)}$ has non-zero term only when $ p = \hat{p}, \phi \neq \hat{\phi}, j= \hat{j}$, and is equal to $\gamma_{{\phi}, \hat{p}}^{(\hat{j})}$. This yields $\Phi-1$ terms:
\vspace{-\EqVspace}
\begin{equation}
\vspace{-\EqVspace}
\ddagger^{(\hat{j})}_{\hat{\phi}, \hat{p}}
(\mathbf{X}, \mathbf{D})
\triangleq
\sum_{{\phi =1, \; \phi \neq \hat{\phi}}}^{\Phi}
W_{\mathrm{b}}
\frac{
-
(
w_{{\phi}} \mathfrak{s}_{\phi}(r_{\phi})
+ s_{{\phi}}
)
\gamma^{(\hat{j})}_{{\phi},\hat{p}}
x^{(\hat{j})}_{{\phi},\hat{p}}
\gamma^{(\hat{j})}_{{\phi},\hat{p}}
}
{
\big(
\gamma_{\mathrm{n}}
+
\mathrm{I}_{\phi, \hat{p}}^{(\hat{j})}
+
\gamma_{{\phi},\hat{p}}^{(\hat{j})}
x_{{\phi},\hat{p}}^{(\hat{j})}
\big)
\big(
\gamma_{\mathrm{n}}
+
\mathrm{I}_{{\phi},\hat{p}}^{(\hat{j})}
\big)
}.
\label{eq:DaggerTwo}
\end{equation}
It is insightful to note that the terms $\dagger^{(\hat{j})}_{\hat{\phi}, \hat{p}}$ and $\ddagger^{(\hat{j})}_{\hat{\phi}, \hat{p}}$ are the \textit{aggregate rate gain sensitivity (for the inter- and intra-cell interference)}, as functions of $x_{\phi, p}^{(j)}$. Dynamic interference coordination occurs when $\Upsilon^{(\hat{j})}_{\hat{\phi}, \hat{p}} + \dagger^{(\hat{j})}_{\hat{p}} + \ddagger^{(\hat{j})}_{\hat{\phi}, \hat{p}} - u_{\hat{p}}^{(\hat{j})} - v^{(\hat{j})}_{\hat{\phi}} = 0$ is satisfied. Intuitively, this is when the \textit{aggregate rate gain in using an RB, minus its dual cost, is equal to the aggregate rate loss in using that RB}.

Having calculated the derivative of the Lagrangian, vanishing the derivative yields to
$\Upsilon^{(\hat{j})}_{\hat{\phi}, \hat{p}}
+
\dagger^{(\hat{j})}_{\hat{p}}
+
\ddagger^{(\hat{j})}_{\hat{\phi}, \hat{p}}
-
u_{\hat{p}}^{(\hat{j})}
-
v^{(\hat{j})}_{\hat{\phi}} = 0$. This results in our novel primal update of
\vspace{-\EqVspace}
\begin{equation}
\vspace{-\EqVspace}
\mathfrak{P}_{\hat{\phi},\hat{p}}^{(\hat{j})}
\big(
\mathbf{X}^{(\hat{j})},
\mathbf{D}
\big)
\triangleq
\Bigg(
\frac{
(
w_{\hat{\phi}} \mathfrak{s}_{\hat{\phi}}( r_{\hat{\phi}} )
+ s_{\hat{\phi}}
)
W_{\mathrm{b}}
}
{
-
\dagger^{(\hat{j})}_{\hat{p}}
-
\ddagger^{(\hat{j})}_{\hat{\phi}, \hat{p}}
+
u_{\hat{p}}^{(\hat{j})}
+
v^{(\hat{j})}_{\hat{\phi}}
}
-
\frac{
\gamma_{\mathrm{n}}
+
\mathrm{I}_{\hat{\phi}, \hat{p}}^{(\hat{j})}
}
{\gamma_{\hat{\phi}, \hat{p}}^{(\hat{j})} }
\Bigg)^{+}
\label{eq:MainPrimal}
\end{equation}
which updates the primal variable $x_{\hat{\phi},\hat{p}}^{(\hat{j})}$.
Here, $(\cdot)^{+} = \max (0, \cdot)$. Importantly, the allocations depend not only on interference but also on the relative priorities due to QoS requirements.

The system of equations in~\eqref{eq:MainPrimal} represents $\Phi P J$ equations, with $\Phi P J$ primal unknowns, and $J (P + \Phi) + \Phi$ dual unknowns. We note that these primal equations have a special structure that each primal unknown can be written \textit{explicitly} in terms of other primal unknowns. This special structure makes the system amenable to fixed point iterations. For this \emph{inner loop}, the dual variables are constants (to be determined in an \emph{outer loop}, in Section~\ref{sec:DualDomain}).

We use $i^{\mathrm{outer}}$ and $i^{\mathrm{inner}}$ to denote the outer and inner loop counters, respectively. In the fixed point method, solving for the primal variables involves iterations (on $i^{\mathrm{inner}}$), for a fixed $i^{\mathrm{outer}}$, as
\vspace{-\EqVspace}
\vspace{-\EqVspace}
\begin{multline}
x_{\hat{\phi},\hat{p}}^{(\hat{j})} [k, i^{\mathrm{outer}}, i^{\mathrm{inner}} + 1]
\\
\leftarrow
\mathfrak{P}_{\hat{\phi},\hat{p}}^{(\hat{j})}
\big(
\mathbf{X}^{(\hat{j})} [k, i^{\mathrm{outer}}, i^{\mathrm{inner}} ],
\mathbf{D} [k, i^{\mathrm{outer}}]
\big).
\end{multline}
The inner loop is terminated if
$ \max_{\phi, p, j}
\big|
x_{\phi, p}^{(j)} [k, i^{\mathrm{outer}}, i^{\mathrm{inner}}]
-
x_{\phi, p}^{(j)} [k, i^{\mathrm{outer}}, i^{\mathrm{inner}} + 1]
\big|
\leq
\epsilon^\mathrm{inner}$, where $\epsilon^\mathrm{inner}$ is the inner loop error tolerance.
A convergence analysis of the associated fixed point method is beyond the scope of this paper.
Having solved the primal equations, the outer loop index is increased passing the results of the primal variable to the next outer iteration as
$\mathbf{X} [k, i^{\mathrm{outer}} +1, 1]
\leftarrow
\mathbf{X} [k, i^{\mathrm{outer}}, {i^{\mathrm{inner}}}^*]
$, where ${i^{\mathrm{inner}}}^*$ is the smallest index satisfying the error.

\vspace{-\BeforeSectionTitle}
\subsection{Dealing with Dual Variables}
\label{sec:DualDomain}
\vspace{-\AfterSectionTitle}
In this section, we solve the outer optimization, $\min_{0\leq\mathbf{D}} \mathfrak{L} (\mathbf{X}, \mathbf{D})$, finding the dual variables for fixed primal variables. Since the value of the Lagrangian is an upper bound on the original optimization and any feasible primal solution provides a lower bound, if the gap $g [k, i^{\mathrm{outer}}] \triangleq |\mathfrak{L} (\mathbf{X} [k, i^{\mathrm{outer}}], \mathbf{D} [k, i^{\mathrm{outer}}]) - \mathfrak{U}_{\mathrm{T}} (\mathbf{X} [k, i^{\mathrm{outer}}]) | \leq \epsilon^{\mathrm{outer}}$, then the primal and dual solutions are within $\epsilon^{\mathrm{outer}}$ of local optimality~\cite{BertsekasCons}. The outer iterations aim at $g [k, i^{\mathrm{outer}} + 1] \leq g [k, i^{\mathrm{outer}}]$, for large $i^{\mathrm{outer}}$. To find the best upper bound, the Lagrange multipliers are updated in the direction opposite to the gradient of the Lagrangian (with respect to the multipliers), in the outer loop.

The bisection method~\cite{ZheAdve} is the standard approach to dual updates. However, in our problem, the several dual variables have very different roles making bisection ineffective. We, instead, design a customized update rule building on the basic idea of \emph{increase (decrease) the multiplier, if the corresponding constraint is violated (satisfied)}. In addition, a computationally efficient update rule should have four features: (i) providing exponential convergence; (ii) accounting for the violation/satisfaction margins; (iii) being able to sweep the whole interval from zero to the multipliers maximum; and finally (iv) shrinking the steps sizes with the outer iteration. The bisection method lacks features (ii) and (iii), making it ineffective in our problem. The classical gradient method lacks feature (i), making it very slow.

We introduce a novel dual update method, based on Lagrangian gradients, combining aforementioned features. Define the Lagrangian gradient with respect to $s_{\phi}$ as $\Delta^{s_{\phi}} \triangleq r_{\phi} [k, i^{\mathrm{outer}}] - r^{\mathrm{min}}_{\phi} [k]$; the Lagrangian gradient with respect to $u^{(j)}_{p}$ as $\Delta^{{u}_{p}^{(j)}} \triangleq 1 - \Sigma_{\phi=1}^{\Phi} x_{\phi, p}^{(j)} [k, i^{\mathrm{outer}}]$; and the Lagrangian gradient with respect to $v^{(j)}_{\phi}$ as $\Delta^{v_{\phi}^{(j)}} \triangleq 1 - \Sigma_{p=1}^{P} x_{\phi, p}^{(j)} [k, i^{\mathrm{outer}}]$. Negative values of any $\Delta$ correspond to a constraint violation, while positive values indicate constraint satisfaction.

We use the multiplicative factors of $\aleph > 1$, if a constraint is violated (and $1/\aleph$ if satisfied). The multiplicative factors provide feature (i). In addition, we amplify them, based on the ratio of satisfaction or violation, based on the mapping $\Delta^{s_{\phi}} \rightarrow \min( |\log (1+ {\Delta^{s_{\phi}}}/{r^{\mathrm{min}}_{\phi}}) |, \delta^{\mathrm{max}})$, in order to provide feature (ii). The image of this mapping is $[0, \delta^{\mathrm{max}}]$, for both cases of violation (${r_{\phi}}/{r^{\mathrm{min}}_{\phi}} \in [0, 1)$) and satisfaction (${r_{\phi}}/{r^{\mathrm{min}}_{\phi}} \in [1, \infty)$) of the minimum rate constraint. Nevertheless, using the term $|\log(\cdot)|$ makes the amplification \emph{aggressive} (acting as an increasing \textit{convex} function), when the constraint is violated, in comparison to when it is satisfied (acting as an increasing \textit{concave} function).
Note the absolute sign in $|\log(\cdot)|$. 
We also use the outer counter to shrink dual update steps, based on dividing the $\min( |\log (1+ {\Delta^{s_{\phi}}}/{r^{\mathrm{min}}_{\phi}}) |, \delta^{\mathrm{max}})$ by ${(i^{\mathrm{outer}})}^{\varpi}$ providing feature (iv). The parameter $\varpi$ controls the intensity of this feature. Importantly, unlike bisection, our method provides feature (iii). The resultant dual update, for variable $s_{\phi}$, is
\begin{multline}
s_{\phi} [k, i^{\mathrm{outer}}+1 ]
\quad \quad \leftarrow
\\
\begin{cases}
\min
\Big(
\lambda^{\mathrm{max}},
\aleph^{
\frac{
\min(
| \log (
1
+
{\Delta^{s_\phi}}/{r^{\mathrm{min}}_{\phi}}
) |
,
\delta^{\mathrm{max}})
}
{{(i^{\mathrm{outer}})}^{\varpi}
}
}
s_{\phi} [k,i^{\mathrm{outer}}]
+
\vartheta
\Big),
& \\
\text{\quad \quad \quad \quad \quad \quad \quad \quad \quad \quad \quad \quad \quad \quad if}
\quad
\Delta^{s_\phi} < 0,
\\
\\
\min \Big(
\lambda^{\mathrm{max}},
\left(\frac{1}{\aleph}\right)^{
\frac{
\min(
| \log (
1
+
{\Delta^{s_\phi}}/{r^{\mathrm{min}}_{\phi}}
) |
,
\delta^{\mathrm{max}})
}
{{(i^{\mathrm{outer}})}^{\varpi}
}
}
s_{\phi} [k,i^{\mathrm{outer}}]
\Big),
& \\
\text{\quad \quad \quad \quad \quad \quad \quad \quad \quad \quad \quad \quad \quad \quad if}
\quad
0 \leq \Delta^{s_\phi},
\end{cases}
\label{eq:DualUpdateS}
\end{multline}
where $\delta^\mathrm{max}$ and $\lambda^\mathrm{max}$ are constants chosen to limit the scale factor in~\eqref{eq:DualUpdateS} and the maximum value of the Lagrangian respectively. These limits help avoid computational issues. The small, positive, bias term ($\vartheta$) \textit{jump starts} the dual variable, when encountering a new violated constraint. We denote the mapping in~\eqref{eq:DualUpdateS} as $s_{\phi} [k, i^{\mathrm{outer}}+1 ] \leftarrow \mathfrak{D}^{\mathrm{s}} (s_{\phi} [k, i^{\mathrm{outer}}], \mathbf{X} [k, i^{\mathrm{outer}}])$.

The dual updates for $u_{p}^{(j)}$ and $v_{\phi}^{(j)}$ are calculated similarly and denoted as $\mathfrak{D}^{\mathrm{u}} (u^{(j)}_{p}, \mathbf{X})$ and $\mathfrak{D}^{\mathrm{v}} (v^{(j)}_{\phi}, \mathbf{X})$, respectively. The overall dual update is denoted by
$\mathbf{D} [k, i^{\mathrm{outer}} + 1]
\leftarrow
\mathfrak{D}
\left(
\mathbf{D} [k, i^{\mathrm{outer}} ],
\mathbf{X} [k, i^{\mathrm{outer}} ]
\right)$.

It is insightful to note that the dual update in~\eqref{eq:DualUpdateS}, together with~\eqref{eq:MainPrimal}, suggests that if a min rate constraint is violated (satisfied), the corresponding multiplier grows (shrinks) and, due to its \emph{positive} effect on~\eqref{eq:MainPrimal}, the use of the corresponding link is more (less) likely. Similarly, if a constraint in $\mathcal{C}^{\mathrm{PHY}_1}$ or $\mathcal{C}^{\mathrm{PHY}_2}$ is violated (satisfied), the corresponding multiplier grows (shrinks), but due to its \emph{negative} effect on~\eqref{eq:MainPrimal}, the use of the corresponding link will be less (more) likely.

\vspace{-\BeforeSectionTitle}
\section{Algorithms: \textsc{QoSaIC}, \textsc{ILM}, and \textsc{Inter-Frame QoSaIC}}
\label{sec:algorithms}
\vspace{-\AfterSectionTitle}
In the previous section, we provided the guideline solving the translated \emph{per-frame} optimization problem in~\eqref{eq:optimVerFinal}: Dual iterations form an outer loop using updates as per~\eqref{eq:DualUpdateS}; Primal iterations form an inner loop updating the primal variables as per~\eqref{eq:MainPrimal}. In this section, we use the guideline, in the previous section, and devise our proposed \textsc{QoS aware Interference Coordination (QoSaIC)} algorithm. We also formulate and describe the novel \textsc{Infeasible Load Management (ILM)} algorithm to deal with infeasible load conditions. Finally, we develop the \textsc{Inter-Frame QoSaIC}, combining \textsc{QoSaIC} with \textsc{ILM} to solve our original problem in~\eqref{eq:optimVer1}.

\vspace{-\BeforeSectionTitle}
\subsection{Algorithm \textsc{QoSaIC}}
\label{sec:AlgQoSaIC}
\vspace{-\AfterSectionTitle}
The algorithm \textsc{QoSaIC} listed below executes the primal-dual iterations. After initialization of primal and dual variables, Steps~$2$-$21$ form the outer loop, over counter $i^{\mathrm{outer}}$ (incremented in Step~$3$). Steps~$4$-$8$ form the inner loop, over counter $i^{\mathrm{inner}}$ (incremented in Step~$5$). Step~$4$ checks for convergence of primal solution and the inner loop timeout. Step~$6$ updates intermediate variables required for executing the fixed point method, in Step~$7$. Step~$9$ projects primal variable into $[0, 1]$, acting as the \emph{primal-dual interface} assisting satisfy the integer constraints~\eqref{eq:constINTVer1} faster.

The inner loop error threshold gradually contracts, in Step~$10$, in order to reduce the computational complexity of the inner loop. It starts from a loose $\epsilon^\mathrm{inner}_{1}$ at the early stages of the outer loop counter and moves towards a tighter $\epsilon^\mathrm{inner}_{\infty}$ ($\epsilon^\mathrm{inner}_{1} \geq \epsilon^\mathrm{inner}_{\infty}$). The intuition is that at the early stages of the outer loop, since the dual variables are far from their optimal values, we can afford higher error threshold on the primal values, in order to reduce complexity. The parameter $\varepsilon > 1$ controls the intensity of the aforementioned contraction. We note that similar approach has been documented in, e.g.,~\cite{YaFengLiu2016}, though we developed this notion independently. Dual variables get updated in Step~$11$.

The primal-dual gap, $g[k, i^{\mathrm{outer}}]$, is calculated in Step~$12$. Ideally, this gap is less than $\epsilon^\mathrm{outer}_1$. Nevertheless, due to the range of different input parameters, it is likely that a fixed gap cannot be satisfied, fast enough, for all frames. Therefore, we gradually increase the threshold on this gap, based on the outer counter (in Step~$13$) by expanding with the factor $\varrho > 1$, after passing each $i^{\mathrm{outer, max}}/5$ iterations. The termination condition (Steps~$15$-$17$) of the outer loop (denoted by binary variable $\mathtt{break}$) is flagged true (Step~$16$), when a feasible solution is found and primal-dual gap is lower than threshold defined in Step~$13$. When the outer iteration counter increases, the algorithm deemphasizes the primal-dual gap guarantee. If the outer iteration counter passes half of its preset maximum ($i^{\mathrm{outer, max}}$), the algorithm gives up on a primal-dual gap guarantee and aims exclusively at feasibility (Steps~$18$-$20$).
We use $[k, i^{\mathrm{outer}}, i^{\mathrm{inner}}]$ to refer to an iteration in inner loop with counter $i^{\mathrm{inner}}$, inside outer loop with counter $i^{\mathrm{outer}}$, and inside frame $k$.

The algorithm complexity is
$ \mathcal{O}
\big(
i^{\mathrm{outer, max}}
(
i^{\mathrm{inner, max}} P (J + \Phi( J + P \Phi ) + \Phi(J + \Phi )$ $+ \Phi )
(8 \Phi P J + 2\Phi )
) \big)$.
Due to space limitations, we do not include the derivation. Nevertheless, we highlight that the complexity is polynomial in all dimensions. Note that \textsc{QoSaIC} is executed for a single specific frame; later, it will become a building block of \textsc{Inter-Frame QoSaIC} algorithm.

\begin{algorithm}[h!]
\label{alg:QoSaIC}
\renewcommand{\thealgorithm}{}
\caption{
\textcolor{black}{
\textsc{QoSaIC}
${ \big(
k,
\boldsymbol{\Gamma} [k], \mathbf{r}^{\mathrm{min}} [k], \mathbf{r}^{\mathrm{max}} [k], \Phi, P, J, T
\big) } \quad \quad \quad$
$\quad \quad \quad \rightarrow
(
\mathbf{X} [k], \mathbf{D} [k]
) $.
}}
\begin{algorithmic} [1]
\State
initialize
$\mathbf{X} [k,1] \; \text{\&} \; \mathbf{D} [k,1]$, $i^{\mathrm{outer}} \leftarrow 0$
\While{
{
$\neg$
$\mathtt{break}$
\textbf{\&}
$i^{\mathrm{outer}} \leq i^{\mathrm{outer, max}}$
}}
\State $i^{\mathrm{outer}} \leftarrow i^{\mathrm{outer}} +1$
\While{
\vspace{-0.65cm}
\begin{align*}
\vspace{-0.9cm}
&\hspace{+0.3cm} {\max}_{\phi, p, j}
\big|
x_{\phi, p}^{(j)} [k, i^{\mathrm{outer}}, i^{\mathrm{inner}}-1]
\\
&-
x_{\phi, p}^{(j)} [k, i^{\mathrm{outer}}, i^{\mathrm{inner}} ]
\big|
\geq
\epsilon^\mathrm{inner} [i^\mathrm{outer}]
\\
&\textbf{\; \& \;}
i^{\mathrm{inner}} \leq i^{\mathrm{inner, max}}
\end{align*}
\hspace{0.5cm}
}
\State $i^{\mathrm{inner}} \leftarrow i^{\mathrm{inner}} +1$
\State
\text{update}
$\dagger_{p}^{(j)}, \dagger_{\phi, p}^{(j)}, \mathrm{I}_{\phi, p}^{(j)}$
\text{based on},~\eqref{eq:DaggerOne}, and~\eqref{eq:DaggerTwo}
\State \vspace{-0.5cm}
\begin{align*}
\vspace{-0.8cm}
&x_{{\phi},{p}}^{({j})}
[k, i^{\mathrm{outer}}, i^{\mathrm{inner}} ]
\leftarrow \\
&\mathfrak{P}_{{\phi},{p}}^{({j})}
\big(
\mathbf{X}^{({j})} [k, i^{\mathrm{outer}}, i^{\mathrm{inner}} -1 ],
\mathbf{D} [k, i^{\mathrm{outer}}]
\big)
\end{align*}
\EndWhile
\State
$\hspace{-0.4cm} x_{\phi, p}^{(j)} [k, i^{\mathrm{outer}}, i^{\mathrm{inner}} ]
\leftarrow
\min
\big(
1,
x_{\phi, p}^{(j)} [k, i^{\mathrm{outer}}, i^{\mathrm{inner}} - 1]
\big)
$
\State
$\epsilon^\mathrm{inner} [i^\mathrm{outer} + 1]
\leftarrow
\epsilon^\mathrm{inner}_\infty
+
(\epsilon^\mathrm{inner}_1 - \epsilon^\mathrm{inner}_\infty) (\varepsilon)^ {1-i^\mathrm{outer}}$
\State
$\mathbf{D} [k, i^{\mathrm{outer}} ]
\leftarrow
\mathfrak{D}
(
\mathbf{D} [k, i^{\mathrm{outer}} -1],
\mathbf{X} [k, i^{\mathrm{outer}} -1]
)
$
\State
\vspace{-0.6cm}
\begin{align*}
\vspace{-0.3cm}
&g[k, i^{\mathrm{outer}}]
\leftarrow \\
&\big|
\mathfrak{L} (\mathbf{X} [k, i^{\mathrm{outer}}], \mathbf{D} [k, i^{\mathrm{outer}}])
-
\mathfrak{U}_{\mathrm{T}}(\mathbf{X} [k, i^{\mathrm{outer}}])
\big|
\end{align*}
\State
$\epsilon^\mathrm{outer} [i^\mathrm{outer}]
\leftarrow
\epsilon^\mathrm{outer}_1
(\varrho)^{
\big\lfloor
\frac{i^{\mathrm{outer}}}{(i^{\mathrm{outer, max}}/5)}
\big\rfloor
}
$
\State $\mathtt{break} \leftarrow \text{false}$
\If
{
\vspace{-0.7cm}
\begin{align*}
\vspace{-0.5cm}
&g[k, i^{\mathrm{outer}}]
\leq
\epsilon^\mathrm{outer} [i^{\mathrm{outer}}]
\\
&\textbf{\&}
\mathbf{X} [k, i^{\mathrm{outer}}, i^{\mathrm{inner}}]
\in
\mathcal{C}^{\mathrm{INT}} \cap \mathcal{C}^{\mathrm{PHY}}
\\
&
\textbf{\&}
\mathbf{r} [k]
\in
\mathcal{C}^{\mathrm{f-MAC}} [k]
\end{align*}
\hspace{+0.4cm}
}
\State
$\mathtt{break} \leftarrow \text{true}$
\EndIf
\If
{\vspace{-0.5cm}
\begin{align*}
\vspace{-0.7cm}
&i^{\mathrm{outer}}
\geq
i^{\mathrm{outer, max}} / 2
\;
\textbf{\&}
\;
\mathbf{r} [k]
\in
\mathcal{C}^{\mathrm{f-MAC}} [k]
\;
\\
&
\;
\textbf{\&}
\;
\mathbf{X} [k, i^{\mathrm{outer}}, i^{\mathrm{inner}}]
\in
\mathcal{C}^{\mathrm{INT}} \cap \mathcal{C}^{\mathrm{PHY}}
\end{align*}
\hspace{0.35cm}
}
\State
$\mathtt{break} \leftarrow \text{true}$
\EndIf
\EndWhile
\end{algorithmic}
\end{algorithm}

\vspace{-\BeforeSectionTitle}
\subsection{QoS Outages: Causes and Quantification}
\label{sec:Outage}
\vspace{-\AfterSectionTitle}
The combination of QoS requirements, random arrivals, and channel scenario determine the feasible set to the frame-level problem (solved by \textsc{QoSaIC}). For a fixed channel and QoS requirements~\footnote{fixed delay targets, fixed number of RS and DS flows, and fixed distribution of overall load among flows}, when we increase the \textit{total mean input rate}, the system passes a load point, where the problem become \textit{infeasible} producing inevitable outage. This is the first cause of outages which occurs in a scenario with high load. In such an \textit{overload} condition, the demands~\footnote{The demand is a combination of QoS requirements and users' traffic (spatial and temporal distributions).} exceed the supply of effective capacity~\footnote{Effective capacity is a function of the bandwidth, channel strengths and the spatial interference scenario.} for most frames. Another cause of outages is when the granularity of the RBs is not fine enough in comparison to the frame rates constraints, i.e., some flows are over-provisioned. This can occur in both low and high load scenarios. The problem may also become infeasible if the arrival rate \textit{spikes}, even if the mean input rate is low. In summary, it is possible that the optimization problem become infeasible (with respect to the translated min frame rates). In such a condition, we must explore reasonable compromises (relaxations) of the QoS constraints to obtain a useful solution (without violating the PHY layer constraints).

Toward finding a systematic compromise, we first define the outages as
\vspace{-\EqVspace}
\begin{multline}
\vspace{-\EqVspace}
O^{\mathrm{r^{min}}}_{\phi} [k]
\triangleq
\Big(
1
-
\frac{r_{\phi}[k]}{r_{\phi}^{\mathrm{min}} [k]}
\Big)^{+},
\;
\\
O^{\mathrm{\bar{r}^{min}}}_{\phi} [k]
\triangleq
\Big(
1
-
\frac{\bar{r}_{\phi}[k]}{\bar{r}_{\phi}^{\mathrm{min}} }
\Big)^{+},
\;
\\
O^{\mathrm{\bar{d}^{max}}}_{\phi} [k]
\triangleq
\Big(
\frac{\bar{d}_{\phi}[k]}{ \bar{d}_{\phi}^{\mathrm{max}} }
-
1
\Big)^{+},
\label{eq:OutageDefinitions}
\end{multline}
where $O^{\mathrm{r^{min}}}_{\phi} [k]$,
$O^{\mathrm{\bar{r}^{min}}}_{\phi} [k]$,
and $O^{\mathrm{\bar{d}^{max}}}_{\phi} [k]$
quantify violations from $r_{\phi}^{\mathrm{min}} [k]$,
$\bar{r}_{\phi}^{\mathrm{min}}$,
and $\bar{d}_{\phi}^{\mathrm{max}}$,
respectively. As metrics for a sequence of frames, we also define the mean outages over frames. For example, the mean outage from mean-delay target, until frame $k$, is defined as
$\bar{O}^{\mathrm{\bar{d}^{max}}}_{\phi} [k]
=
\frac{1}{k}
\sum_{k'=1}^{k}
O^{\mathrm{\bar{d}^{max}}}_{\phi} [k'] $.
Having introduced the outages, we also define the \emph{underloaded} condition, if all the mean outages are zero; otherwise, we say the system is in the \emph{overloaded} condition. The condition depends on the combination of channels, QoS requirements, and arrival processes.

\vspace{-\BeforeSectionTitle}
\subsection{The Infeasible Load Manager (\textsc{ILM}) Algorithm}
\label{sec:ILM}
\vspace{-\AfterSectionTitle}
The goal of the \textsc{ILM} algorithm is to relax the QoS constraints that are deemed infeasible by the \textsc{QoSaIC} algorithm with a small practical deviation from QoS constraints. To design the compromise approach, we choose to minimize the summation of the squared mean outage (from minimum frame rate). Having an objective, we use its gradient (with respect to $\mathbf{r} [k]$), in order to find which flow to compromise. The corresponding gradient vector is
\vspace{-\EqVspace}
\begin{equation}
\vspace{-\EqVspace}
\nabla_{\mathbf{r} [k]}
\Big(
\Sigma_{\phi=1}^{\Phi}
(
\bar{O}^{\mathrm{r^{min}}}_{\phi} [k]
)^2
\Big)
=
-\frac{2}{k}
\Big(
\frac{
\bar{O}^{\mathrm{r^{min}}}_{\phi} [k]
}{
r_{\phi}^{\mathrm{min}} [k]
},
\cdots,
\frac{
\bar{O}^{\mathrm{r^{min}}}_{\Phi} [k]
}{
r_{\Phi}^{\mathrm{min}} [k]
}
\Big),
\label{eq:DerivativeOfSumSquareOutage}
\end{equation}
where we used the chain rule, the definition of mean outage from min frame rates
($\bar{O}^{\mathrm{r^{min}}}_{\phi} [k] = \frac{1}{k} \sum_{k'=1}^{k} O^{\mathrm{r^{min}}}_{\phi} [k']$), and first relationship from \eqref{eq:OutageDefinitions}, in the derivation. The largest element of this gradient vector corresponds to the element with the smallest absolute value. The gradient in~\eqref{eq:DerivativeOfSumSquareOutage} suggests that, in order to deal with an infeasible condition, we should relax the flow with the highest ratio of the min rate requirement over mean outage. This would have the least impact on the sum squared mean outage. We denote the index of the corresponding flow with $\phi^{\mathrm{cmp-min}}$ forming the core of \textsc{ILM} algorithm:
$\phi^{\mathrm{cmp-min}}
\leftarrow
\arg
\max_{\phi}
\frac{ r_{\phi}^{\mathrm{min}} [k] } { \bar{O}^{\mathrm{r^{min}}}_{\phi} [k] }$.

Algorithm \textsc{ILM} below delineates the QoS relaxation. The algorithm finds $\phi^{\mathrm{cmp-min}}$, in Step~$1$. Since the outage, in frame $k$, is unknown in Step~1, we use an estimate, based on previous frames. The minimum frame requirement of the flow $\phi^{\mathrm{cmp-min}}$ is decreased by a predetermined factor, $\gimel^{\mathrm{dec}} < 1$, in Step~$3$. Since, for a single frame, \textsc{ILM} can be called several times, the algorithm gives up on a flow by turning off its min frame requirement, if the flow has been compromised more than the $\lfloor \log(\gimel^{\mathrm{dec}})/\log(\sigma)\rfloor$ times (the condition in Step~$2$). If ILM is triggered, the compromised QoS (generated by \textsc{ILM} and denoted as $\mathbf{r}^{\mathrm{cmp-min}} [k]$) are fed back to the algorithm \textsc{QoSaIC}, i.e., they replace the original QoS constraints. Ideally, we would eliminate the condition in Step~$2$; however, it is computationally advantageous to keep it, as it allows \textsc{QoSaIC} to generate a feasible solution faster. Interestingly, we note that Step~$1$ is intuitive as it relaxes the requirement of the \textit{costly flow (high min rate requirement) that experienced low outage}, up to the current frame.

We emphasize that \textsc{ILM} is mainly triggered in overloaded conditions. The notation $\tilde{r}^{\mathrm{min}}_{\phi} [k]$ denotes a copy of the \textit{original} frame requirements. Note that $\mathbf{r}^{\mathrm{min}} [k]$ could be coming from a previous compromise; therefore, in general $r^{\mathrm{min}}_{\phi} [k] \neq \tilde{r}^{\mathrm{min}}_{\phi} [k]$. Moreover, there is a tradeoff between approaching the lowest possible outage and the associated computational complexity. A theoretical analysis of the \textsc{ILM} is beyond the scope of this paper.

\label{sec:ILMmin}
\vspace{-\EqVspace}
\begin{algorithm}[]
\label{alg:ILM}
\renewcommand{\thealgorithm}{}
\caption{
\textsc{ILM}
${\small
\big(
\mathbf{r}^{\mathrm{min}} [k],
\mathbf{O}^{\mathrm{r^{min}}} [k-1] \big)
\rightarrow (\mathbf{r}^{\mathrm{cmp-min}} [k]) }$
}
\begin{algorithmic} [1]
\State
$
\phi^{\mathrm{cmp-min}}
\leftarrow
\arg\min_{\phi}
\frac{r^{\mathrm{min}}_{\phi} [k]}
{\bar{O}^{\mathrm{r^{min}}}_{\phi} [k-1]}
$
{$\quad \quad \quad \quad \quad \quad \quad \quad$ \color{\CMTcolor}\Comment{select the flow (high requirement and low outage)}}
\If{ $r^{\mathrm{min}}_{\phi^{\mathrm{cmp-min}}} [k]
\geq
\sigma \tilde{r}^{\mathrm{min}}_{\phi^{\mathrm{cmp-min}}} [k] $}
{$\quad \quad \quad \quad \quad \quad \quad \quad$ \color{\CMTcolor}\Comment{if current requirement $\geq$ $\sigma\; *$ original requirement}}
\State
$
r^{\mathrm{cmp-min}}_{\phi^{\mathrm{cmp-min}}} [k]
\leftarrow
\gimel^{\mathrm{dec}}*r^{\mathrm{min}}_{\phi^{\mathrm{cmp-min}}} [k]
$
{$\quad \quad \quad \quad \quad \quad \quad \quad$ \color{\CMTcolor}\Comment{relax the requirement partially}}
\\
\textbf{else} $\;$
$r^{\mathrm{cmp-min}}_{\phi^{\mathrm{cmp-min}}} [k]
\leftarrow 0$
{\color{\CMTcolor}\Comment{relax the requirement fully}}
\EndIf
\end{algorithmic}
\end{algorithm}

\vspace{-\BeforeSectionTitle}
\subsection{The \textsc{Inter-Frame QoSaIC} Algorithm}
\label{sec:InterFrameQoSaIC}
\vspace{-\AfterSectionTitle}
We are now ready to present the overall algorithm addressing our original question of \textit{finite backlog queue-aware interference aware resource allocation with mean-rate/delay constraints and random arrivals}. In Section~\ref{sec:SysModelProblemFormulation}, we translated these QoS constraints into \emph{per-frame} parameters, based on \textsc{QoSiFT}. The per-frame constraints were augmented by the frugality constraint incorporated into the objective function. This led to a \emph{frame-level} optimization problem solved by algorithm \textsc{QoSaIC} (coupled by algorithm \textsc{ILM}, if required). In this section, we integrate these algorithms into \textsc{Inter-Frame QoSaIC}, leading to resource allocation \textit{across frames}.

Algorithm \textsc{Inter-Frame QoSaIC} creates a \emph{frame loop} around \textsc{QoSaIC}, in Steps~$1$-$14$. Here, $\mathbf{a}[k]$ denotes the vector of arrivals to the flows making it possible to have finite backlog random arrivals (random arrivals combined with finite backlog is one of the important distinctions of our work). Steps~2 and~3 perform the \emph{arrival update} and \emph{requirements update} (calling \textsc{QoSiFT} from Section~\ref{sec:RefiningTheObjective}), respectively. Steps~$4$-$11$ repeatedly call \textsc{QoSaIC} and \textsc{ILM}, until the compromised requirements become feasible to the core optimization. These steps begin by calling \textsc{QoSaIC} in Step~$5$ while setting a flag variable $\mathtt{trig}$ to false, in Step~$6$. If \textsc{QoSaIC} does not return a feasible solution (Step~$7$), the algorithm calls \textsc{ILM}, in Step~$8$, compromises requirements, in Step~$9$, and makes the $\mathtt{trig}$ true (to return to Step~$5$). When the compromiser is triggered (which is signalled by $\mathtt{trig}$), it returns $\mathbf{r}^{\mathrm{cmp-min}} [k]$.

\begin{algorithm}[h]
\label{alg:IFQoSaIC}
\renewcommand{\thealgorithm}{}
\caption{
\textsc{Inter-Frame QoSaIC}
${\small \big(
\{ \boldsymbol{\Gamma} [k] \}_{1}^{K},
\{ \mathbf{a} [k] \}_{1}^{K},
\bar{\mathbf{r}}^{\mathrm{min}},
\bar{\mathbf{r}}^{\mathrm{max}},
\bar{\mathbf{d}}^{\mathrm{max}},
\Phi, P, J, T \big)
\rightarrow
\big( \bar{\mathbf{r}} [k], \bar{\mathbf{d}} [k] \big) }$
}
\begin{algorithmic} [1]
\For{ $k = 1$ \textbf{to} $k=K$ }
{ \color{\CMTcolor}\Comment{frame loop}}
\State $\mathbf{q} [k] \leftarrow \mathbf{q} [k] + \mathbf{a} [k]$
{ \color{\CMTcolor}\Comment{arrival update}}
\State
\textsc{QoSiFT}
$\small (
{q}_{\phi} [k-1],
{r}_{\phi} [k-1],
\bar{r}_{\phi}^{\mathrm{min}},
\bar{r}_{\phi}^{\mathrm{max}},
\bar{d}_{\phi}^{\mathrm{max}}
)
\rightarrow
(
r_{\phi}^{\mathrm{min}} [k],
r_{\phi}^{\mathrm{max}} [k],
w_{\phi} [k]
)
,
\; \mathtt{trig} \leftarrow \text{true}$
{ \color{\CMTcolor}\Comment{requirements update}}
\While{ $\mathtt{trig}$}
{ \color{\CMTcolor}\Comment{loop, for calling \textsc{QoSaIC}, and if necessary \textsc{ILM}}}
\State
\textsc{QoSaIC}
$\big(
k,
\boldsymbol{\Gamma} [k], \mathbf{r}^{\mathrm{min}} [k], \mathbf{r}^{\mathrm{max}} [k], \Phi, P, J, T
\big)
\rightarrow
(
\mathbf{X} [k], \mathbf{D} [k]
)$
{ \color{\CMTcolor}\Comment{call \textsc{QoSaIC}}}
\State $\; \mathtt{trig} \leftarrow \text{false}$
{ \color{\CMTcolor}\Comment{toggle the while condition}}
\If{
$\neg($
$\mathbf{X} [k] \in \mathcal{C}^{\mathrm{PHY}}$
\textbf{\&}
$\mathbf{r} [k] \in \mathcal{C}^{\mathrm{f-MAC}}$
$)$}
{ \color{\CMTcolor}\Comment{if not fully feasible, call \textsc{ILM}}}
\State
\textsc{ILM
$(\mathbf{r}^{\mathrm{min}} [k], \mathbf{O}^{\mathrm{r^{min}}} [k-1]) \rightarrow (\mathbf{r}^{\mathrm{cmp-min}} [k])$
}
{ \color{\CMTcolor}\Comment{call \textsc{ILM}}}
\State $\mathbf{r}^{\mathrm{min}} [k] \leftarrow \mathbf{r}^{\mathrm{cmp-min}} [k]$, $\mathtt{trig} \leftarrow \text{true}$
{ \color{\CMTcolor}\Comment{substitute the new requirements}}
\EndIf
\EndWhile
\State $\mathbf{q} [k] \leftarrow ( \mathbf{q} [k] - T_{\mathrm{b}} \mathbf{r} [k])^{+}$,
$\mathbf{r} [k] \leftarrow \min(\mathbf{r} [k] , \mathbf{q} [k]/ T_{\mathrm{b}})$
{ \color{\CMTcolor}\Comment{service update}}
\State \text{update} $\bar{\mathbf{q}} [k], \bar{\mathbf{r}} [k], \bar{\mathbf{d}} [k]$
{ \color{\CMTcolor}\Comment{mean quantities update}}
\EndFor
\end{algorithmic}
\end{algorithm}

Step~$12$ executes the \emph{service update} which includes emptying queues on the determined subchannels, through the determined APs, based on $\mathbf{r} [k]$. If $\mathbf{r} [k]$ is larger than the backlog, it is clipped to the backlog amount, as in $\mathbf{r} [k] \leftarrow \min( \mathbf{r} [k], \mathbf{q} [k]/ T_{\mathrm{b}})$. Note that both $\mathbf{q} [k]$ and $\mathbf{a} [k]$ are in bits, while $\mathbf{r}$ is in bps/Hz (corresponding to $W_{\mathrm{b}} = 1$ and $T_{\mathrm{b}} = 1 $), therefore, $T_{\mathrm{b}}$ is reflected in the frugality constraint, service update, and requirements update. Step~$13$ updates the mean values, preparing for requirements update, Step~$3$, in the next frame.

The proposed approach in Section~\ref{sec:Solution} together with several subtle techniques in developing \textsc{QoSaIC} and \textsc{ILM} algorithms forms our third and fourth major contributions (Section~\ref{sec:Contributions}).

\vspace{-\BeforeSectionTitle}
\section{Numerical Examples}
\label{sec:simulations}
\vspace{-\AfterSectionTitle}
\subsection{Scenario and Evaluation}
\label{sec:ParameterValues}
\vspace{-\AfterSectionTitle}
This section presents results of simulations illustrating the efficacy of the algorithms developed. The results of the \textsc{Inter-Frame QoSaIC} algorithm are compared to that of the baseline PF approach. The simulations use $\Phi = 8$ flows corresponding to $\Phi$ randomly distributed users within a 800m $\times$ 800m square, served by $P = 4$ APs using $J = 5$ frequency RBs. Each RB spans $W_{\mathrm{b}} = 180$kHz and is allocated for $T_{\mathrm{b}} = 1$ms. Links between users and APs are modeled with a pathloss exponent of $3.5$. The parameters, used in the algorithms are $\nu = 0.1$ (Sigmoid parameter), $\epsilon^\mathrm{inner}_1 = 10^{-1}$, $\epsilon^\mathrm{inner}_\infty = 10^{-2}$, $i^{\mathrm{inner, max}} = 30$, $\varrho = 1.05$ (inner loop parameters), $\sigma = 10^{-3}$, $\gimel^{\mathrm{dec}} = 0.6$ (\textsc{ILM}), $\aleph = 2$, $\varpi = 0.25$, $\vartheta = 0.1$, $\delta^{\mathrm{max}} = 5$, $\lambda^{\mathrm{max}} = 10^{8}$, $i^{\mathrm{outer, max}} = 150$ (dual update and outer loop parameters). We run the simulations for $K=100$ frames and average over the arrivals.

We categorize the experiments into two groups: single load and multiple load experiments. For the single load, we perform two tests: single load single frame and single load multiple frames. For the multiple load experiments, we test scenarios of (i) having BE and DS flows, (ii) BE, RS and DS flows, and (iii) all types of flows, with emphasis on \textit{heterogeneous arrival rates for the DS flows}. If relevant, DS flows requirements are set to $\bar{d}^{\mathrm{max}}_{\phi} = 20$ frames, in all the experiments. The rate requirements of the RS flow are set based on their input mean; $\bar{r}_{\phi}^{\mathrm{min}} \leftarrow \bar{r}^{\mathrm{in}}_{\phi}$, where $\bar{r}^{\mathrm{in}}_{\phi}$ denotes the mean arrival to flow $\phi$. The arrivals are generated using a Poisson distribution.

We use two crucial metrics to evaluate the algorithms: measuring QoS satisfaction and output rates. The QoS satisfaction is evaluated through the outage definitions in Section~\ref{sec:Outage}: satisfaction of the maximum mean-delay and minimum mean-rate. These metrics evaluate whether the demands are satisfied (zero outage), or if not, measure the \textit{relative distance to demands targets}.

We denote the mean of the input rate, per flow, by $\bar{r}^{\mathrm{in}}_{\phi} = \frac{\bar{a}_{\phi}}{T_{\mathrm{b}}} = \frac{1}{ T_{\mathrm{b}}K}{ \Sigma_{k=1}^{K} a_{\phi}[k]}$, in bps. Similarly, we denote the total mean input rate (over frames, for a single load) by $\bar{r}^{\mathrm{in}}_{\Sigma} = \frac{1}{T_{b}K} \sum_{\phi = 1}^{\Phi} \sum_{k=1}^{K} a_{\phi} [k] $, in bps, where $K$ is number of frames.
The counterparts of the \emph{mean input rates} are the \emph{mean output rates}. As defined in~\eqref{eq:constFramebitrateVer1}, \textit{output frame rates} are $r_{\phi} [k] =\Sigma_{j=1}^{J} \Sigma_{p=1}^{P} \log (1 + \mathsf{SINR}_{\phi,p}^{(j)} [k])$, in bps/Hz. The \textit{mean output rate}, corresponding to flow $\phi$, is $\bar{r}^{\mathrm{out}}_{\phi} \triangleq \frac{1}{K}{\Sigma_{k=1}^{k=K} \min( r_{\phi}[k], q_{\phi}[k]/ T_{\mathrm{b}})} $, in bps/Hz, i.e., output rates are limited by input rates.

\vspace{-\BeforeSectionTitle}
\subsection{Single Load Single Frame Experiment}
\vspace{-\AfterSectionTitle}
\noindent In the first set of results, we set $\mathcal{F}^{\mathrm{DS}} =\{7, 8\}, \mathcal{F}^{\mathrm{RS}} =\{3, 5\}, \mathcal{F}^{\mathrm{BE}} =\{1, 2, 4, 6\}$.

\begin{figure}[]
\centering
\includegraphics[trim={0 9.5cm 0 0},clip, width=1.0\linewidth]{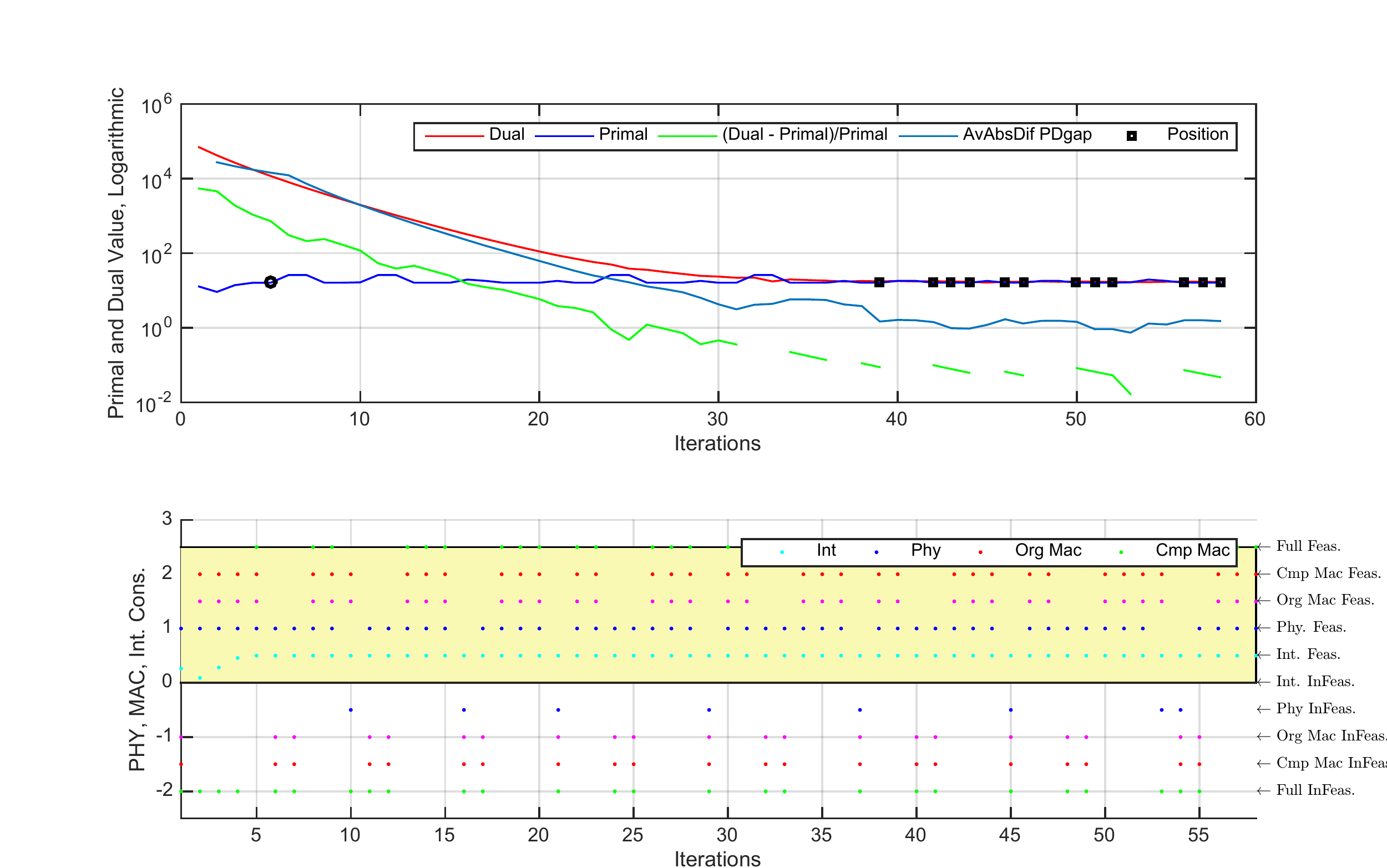} 
\captionsetup{justification=centering}
\vspace{-0.2cm}
\caption{Outer loop convergence, single frame}
\label{fig:Covergence}
\end{figure}

The upper plot in Figure~\ref{fig:Covergence} depicts the primal value, the dual value, the primal-dual gap, and a frame-averaged difference of the primal-dual gap, all versus outer loop iterations. Since we use logarithmic domain, whenever the value of the green curve, (Dual-Primal)/Primal, is negative, the point is absent. We use the fourth curve, in this plot, to check when the primal-dual gap becomes almost a constant. The black dots in the upper plot show that a candidate solution (in a specific outer iteration index) is feasible.

The lower plot decomposes the feasibility depicting the satisfaction/violation for the different constraints. Dots in the lower plot are binary indicators on whether an specific set of constraints is satisfied or not: if a dot is in the shaded region, the corresponding constraint is satisfied, otherwise it is violated. In each outer iteration, the candidate solution is tested against four sets of constraints: INT, PHY, original MAC, and compromised MAC constraints, where the lines of $y = 0.5$, $y = 1.0$, $y = 1.5$, $y = 2.0$ are chosen to represent the satisfaction of the four constraints, respectively (their counterparts in the unshaded region represent a constraint violation). The line $y = 2.5$ corresponds to the case where all the constraints are satisfied. This line is also copied with black dots to the upper plot of the Figure~\ref{fig:Covergence}. We note again that the black dots (in upper plot) represent a logical AND of the satisfaction of all the constraints.

We highlight that if the algorithm \textsc{QoSaIC} cannot find a feasible solution, in less than $i^{\mathrm{outer, max}}$ iterations, we \textit{cannot} choose $\mathbf{X}$ from the last iteration (because it may not be even feasible). Instead, if the outer loop counter gets close to its maximum, we gradually compromise on the primal-dual gap (Step~$13$-$20$ in \textsc{QoSaIC}, page~\pageref{alg:QoSaIC}) while \emph{not} compromising full feasibility. In the case that \textsc{QoSaIC} does not return a fully feasible solution, after $i^{\mathrm{outer, max}}$ iterations, \textsc{ILM} is invoked (see Step~$7$ in \textsc{IFQoSaIC}, page~\pageref{alg:IFQoSaIC}). In this case, we only compromise QoS constraints making the setup robust to the input load scenarios. Feasibility with regards to $\mathcal{C}^{\mathrm{INT}}$, $\mathcal{C}^{\mathrm{{PHY}_1}}$, $\mathcal{C}^{\mathrm{{PHY}_2}}$, in \eqref{eq:optimVer1}, is always strictly preserved. This robustness to the input load is another distinction of this work, in contrast to other works which are fragile to the input load scenarios.

Our extensive simulations show that proposed algorithm typically converges in less than $100$ iterations of the outer loop, similar to the upper plot of Figure~\ref{fig:Covergence}. We note that although, in the early iterations, the algorithm may find a fully feasible solution, it continues to find a better solution, closing the primal-dual gap, if the predefined maximum on outer loop counter allows.

\vspace{-\BeforeSectionTitle}
\subsection{Multiple Loads Experiments}
\vspace{-\AfterSectionTitle}
We categorize the multiple-load experiment into three QoS scenarios of (i) BE+DS flows, (ii) BE+RS+DS flows, and (iii) BE+RS+DS with heterogeneous arrivals. The two metrics of the objective and outages are summarized into scatter plots (for both the proposed scheme and the baseline PF scheme). We also plot the output rates versus input rates, for these three scenarios. To test multiple loads, we use non-uniform points, with a greater concentration near our estimate of the transition from underload to overload, covering this region with higher resolution.

{\label{page:Amended}
Importantly, since simulations are done for \emph{finite number of frames}, at the last frame, there are always residual bits in the queues. Increasing the number of frames can reduce this effect, but the impact remains. To account for these residual bits, we amend the output rates achieved as follows: we use the binary variable $o^{\mathrm{r_{min}}}_{\phi} [k] \in \{0, 1\}$ which indicates outage in a frame, and calculate its mean as $\bar{o}^{\mathrm{r_{min}}}_{\phi} = \frac{1}{K}\Sigma_{k'=1}^{K} o^{\mathrm{r_{min}}}_{\phi} [k']$. Then, we estimate the empirical probability of transmitting (the residual bits), without outage, as $1 - \bar{o}^{\mathrm{r_{min}}}_{\phi}$. Accordingly, we use an \emph{amended output} of $\bar{r}^{\mathrm{out}}_{\phi} + (1 - \bar{o}^{\mathrm{r_{min}}}_{\phi}) r^{\mathrm{res}}_{\phi}$ as a better estimate for the achievable output rates, where $r^{\mathrm{res}}_{\phi} \triangleq q_{\phi} [K] /T_{\mathrm{b}}$.
}

\vspace{-\BeforeSectionTitle}
{\subsubsection{\underline{BE+DS Test}}
\vspace{-\AfterSectionTitle}
In this experiment, $\mathcal{F}^{\mathrm{DS}} =\{7, 8\}$, $\mathcal{F}^{\mathrm{BE}} =\{1, 2, 3, 4, 5, 6\}$, and $\bar{d}_{\phi \in \mathcal{F}^{\mathrm{DS}}}^{\mathrm{max}} = 20$. Figure~\ref{fig:FirstTest_ObjVsOutDS} depicts the scatter plot of the objective (sum-rate on $y$-axis) and outage from the mean-delay requirement ($x$-axis). Each scatter point shows the value of the objective and the corresponding outage. The scatter points are numbered according to their load index, for a total of $12$ input load points (corresponding to the mean input loads in Figure~\ref{fig:FirstTest_ObjVsLoadDS}, from $0$ to $4.5$~Mbps). For both the proposed and baseline PF algorithms, curves are separated into the achieved objective function of all flows and the objective function exclusively for the DS flows.

\begin{figure}[th]
\includegraphics[width=1.00\linewidth]{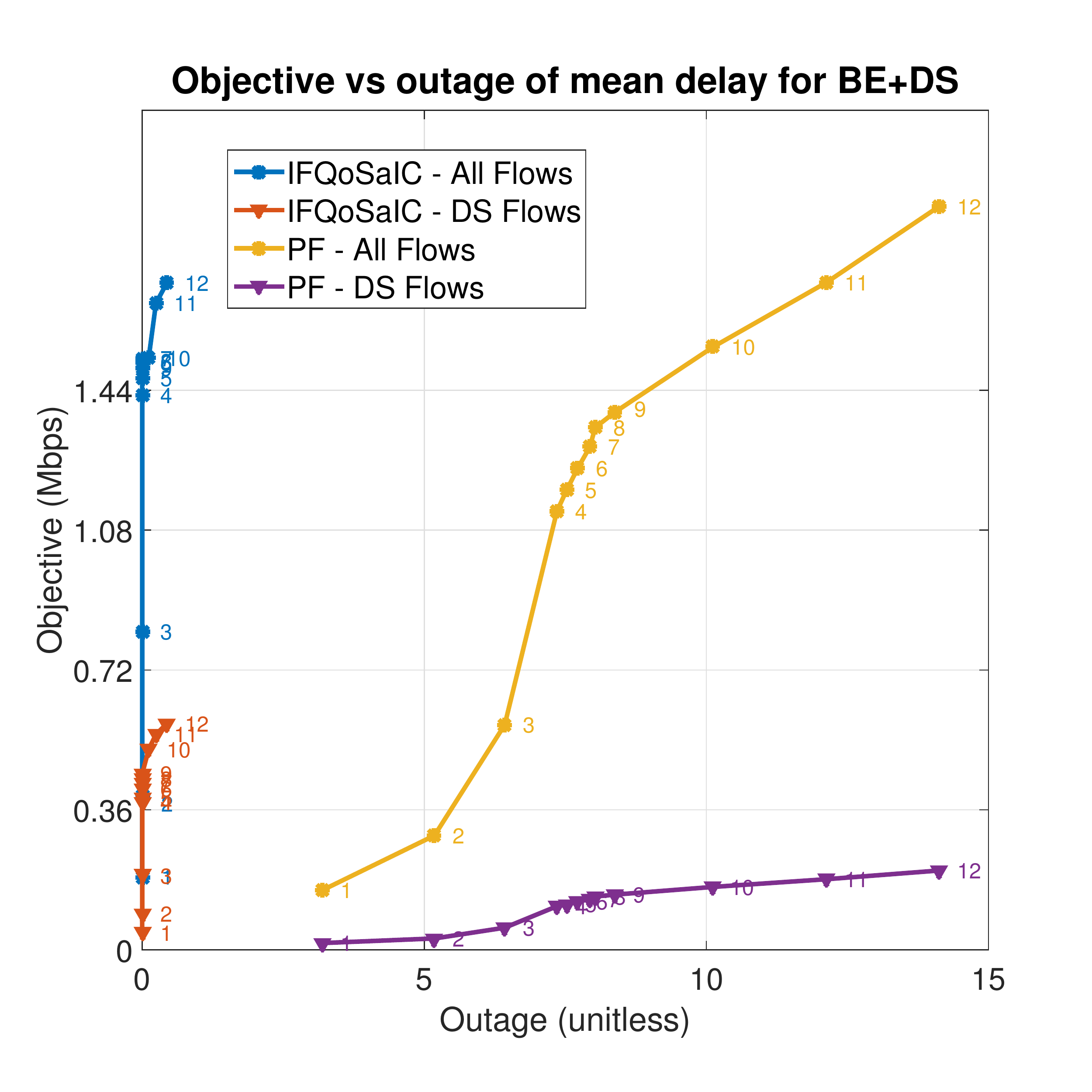}
\captionsetup{justification=centering}
\vspace{-0.4cm}
\caption{Scatter plot, BE+DS experiment}
\label{fig:FirstTest_ObjVsOutDS}
\end{figure}

\begin{figure}[th]
\includegraphics[width=1.00\linewidth]{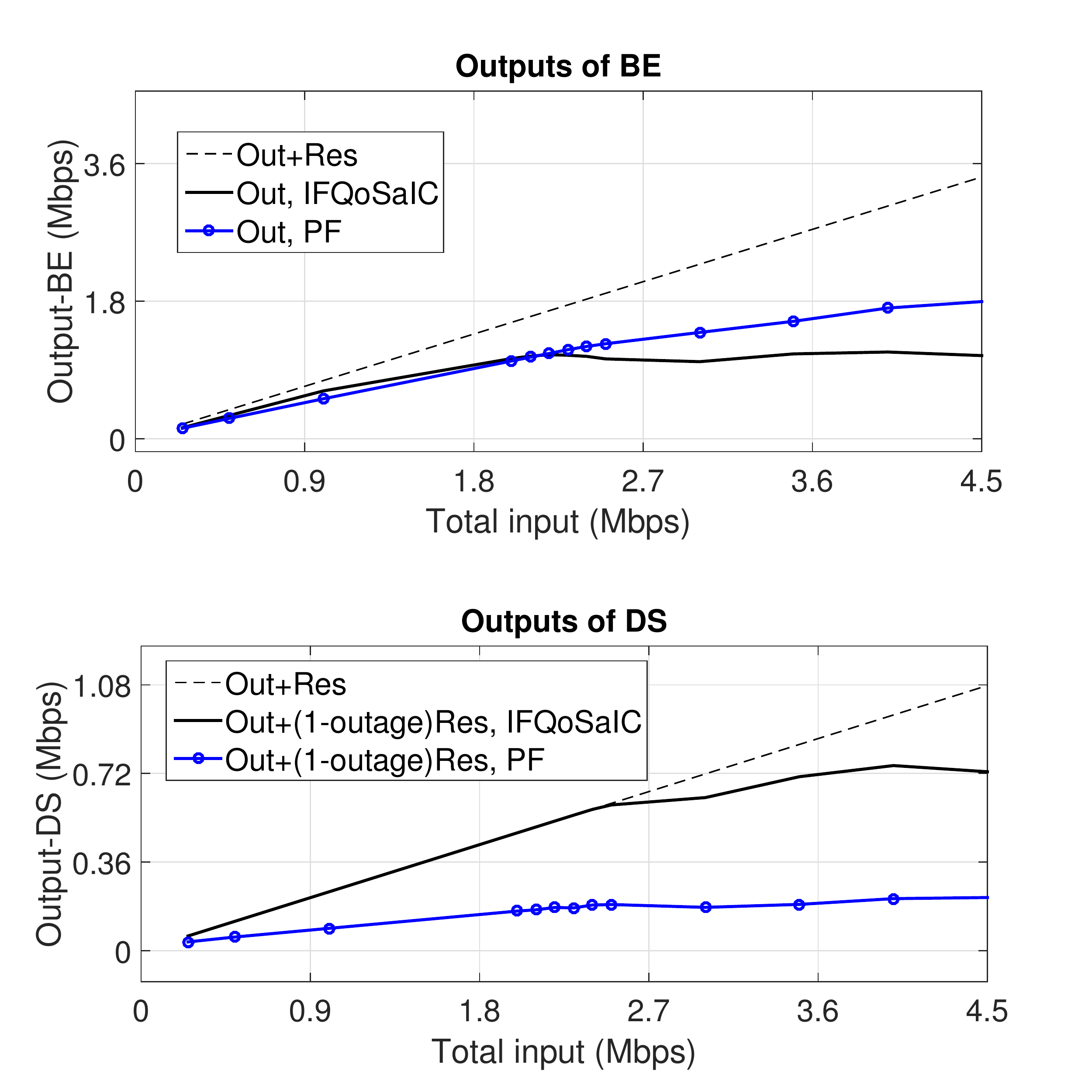}
\captionsetup{justification=centering}
\vspace{-0.4cm}
\caption{Output rates, BE+DS experiment}
\label{fig:FirstTest_ObjVsLoadDS}
\end{figure}

Figure~\ref{fig:FirstTest_ObjVsOutDS} illustrates the clear benefit of the proposed algorithm: maximizing the overall rate while meeting the DS flows constraints. As the figure depicts, the PF leads to extremely high outages. Our algorithm achieves a high overall rate with very low outage (if any). As an example, the rates achieved (say for the input load point $10$, equivalent to $3$~Mbps input rate) by the DS flows is about twice as high as with PF, with effectively negligible outage. The system reaches an overloaded condition around the load point $9$, equivalent to an input rate of $2.5$~Mbps.

For the same experiment, Figure~\ref{fig:FirstTest_ObjVsLoadDS} plots the amended output rates versus input rates for the proposed algorithm (upper) and the baseline PF approach (lower). The lines marked ``Out+Res'' denote the sum of the throughput and the residuals, reflecting the bit conservation law.

\label{sec:StartToBuild}
We observe that, for the BE flows, our algorithms outperform the PF allocation in the underloaded condition (until the input rate of $2.3$~Mbps). From that point, the DS flows get strict priority, and therefore, BE flows suffer. Moreover, we observe that the output of DS flows essentially saturates beyond load number $9$. This is in accordance with the Figure~\ref{fig:FirstTest_ObjVsOutDS}, which indicates that the transition to an overload happened near the same load point. When the amended outputs depart the ``Out+Res'', the queues corresponding to the DS flows start to build up unboundedly - however, the outage rates are far lower than those when using PF.

\vspace{-\BeforeSectionTitle}
\subsubsection{\underline{BE+RS+DS Test}}
\vspace{-\AfterSectionTitle}
In this experiment, the QoS scenario is as $\mathcal{F}^{\mathrm{DS}} =\{7, 8\}$, $\mathcal{F}^{\mathrm{RS}} =\{3, 5\}$, $\mathcal{F}^{\mathrm{BE}} =\{1, 2, 4, 6\}$, with $\bar{d}_{\phi \in \mathcal{F}^{\mathrm{DS}}}^{\mathrm{max}} = 20$. Here, we use two scatter plots: the objective and outage from mean-rate requirements (Figure~\ref{fig:ThirdTest_ObjVsOutDS}), and the objective and outage from mean-delay requirements (Figure~\ref{fig:ThirdTest_ObjVsOutRS}), again comparing to the baseline PF case. As in the earlier test, for a large range of loads, the constraints on delay and rate are met; importantly, our algorithm significantly outperforms the baseline PF approach for DS (Fig.~\ref{fig:ThirdTest_ObjVsOutDS}) and RS (Fig.~\ref{fig:ThirdTest_ObjVsOutRS}) flows. Similar to our previous test, in overloaded conditions (load point $11$ in RS plot and $11-12$ in the DS plot), the total output rates of the baseline are better than for the proposed algorithm. This is because, to limit outages, our algorithm \emph{prioritizes sensitive flows} at the cost of BE flows.


Figure~\ref{fig:ThirdTest_ObjVsLoad} plots the amended output rates versus input rates, for this experiment. The upper figure plots the output of the BE flows. The middle figure plots the amended output for DS flows exclusively. Finally, the lower figure plots the same curves, exclusively for RS flows. We observe that in our algorithm, the outputs of both RS and DS flows outperform those of the baseline, for all loads. These outputs are saturated in overloaded conditions. For all experiments, and especially in this experiment, we observe a small fall in the output of sensitive flows after transiting to non-zero outage. We believe that this fall is the effect of using the \textsc{ILM} algorithm.

\begin{figure}[t!]
\includegraphics[width=1\linewidth]{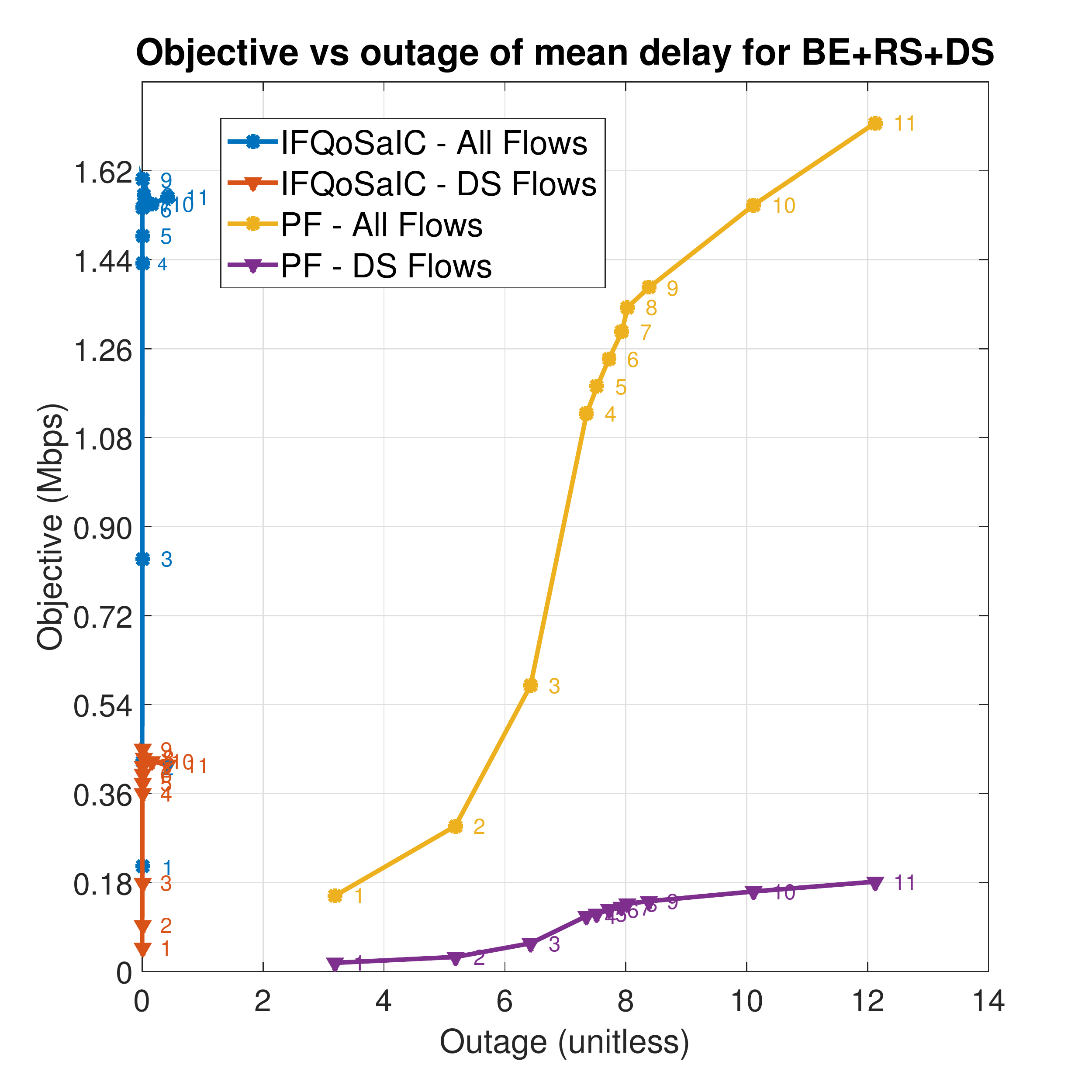}
\captionsetup{justification=centering}
\vspace{-0.4cm}
\caption{Scatter plot of objective and DS outage, BE+DS+RS experiment}
\label{fig:ThirdTest_ObjVsOutDS}
\end{figure}

\begin{figure}[th]
\includegraphics[width=1\linewidth]{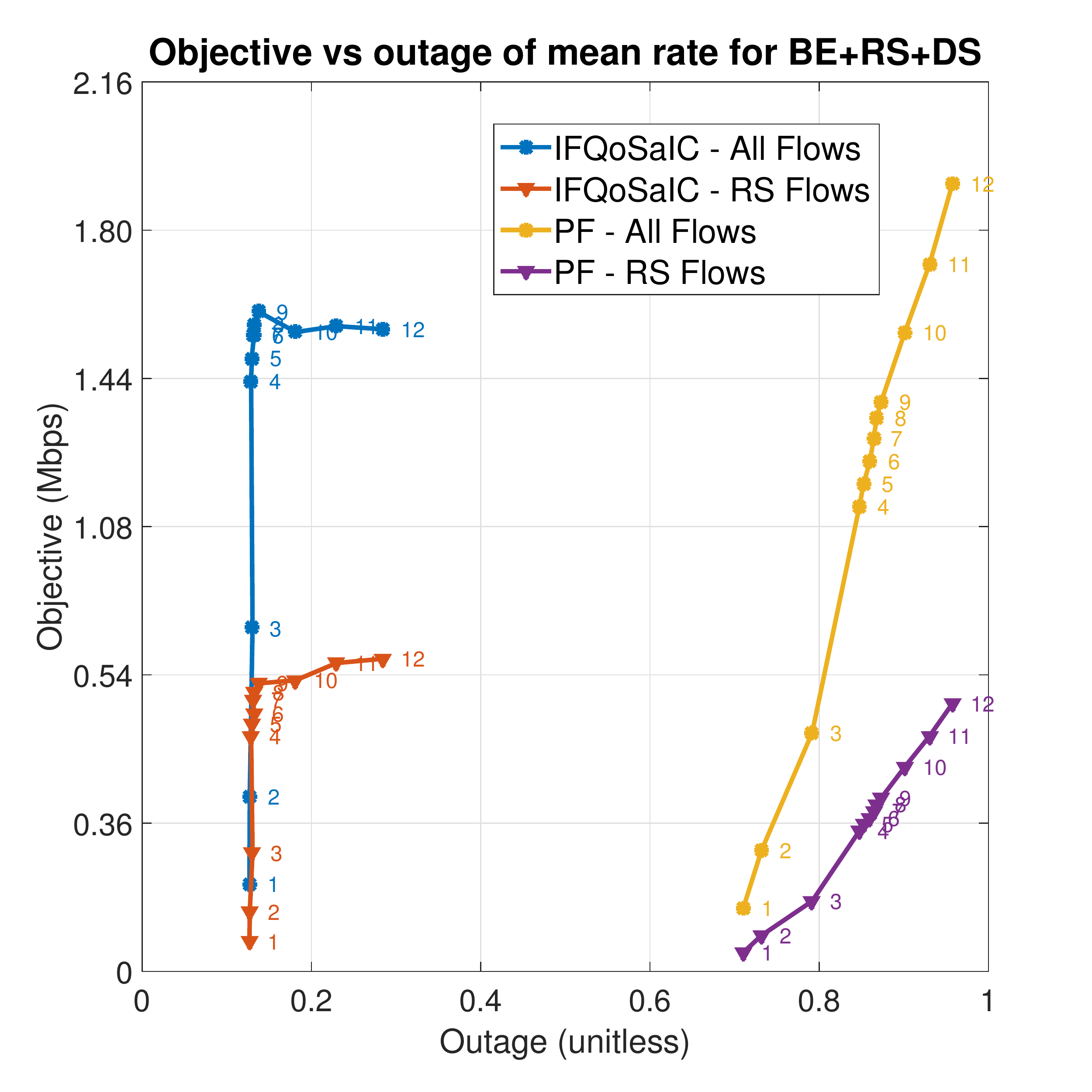}
\captionsetup{justification=centering}
\vspace{-0.4cm}
\caption{Scatter plot of objective and RS outage, BE+DS+RS experiment}
\label{fig:ThirdTest_ObjVsOutRS}
\end{figure}

\begin{figure}[h]
\centering
\includegraphics[width=1\linewidth]{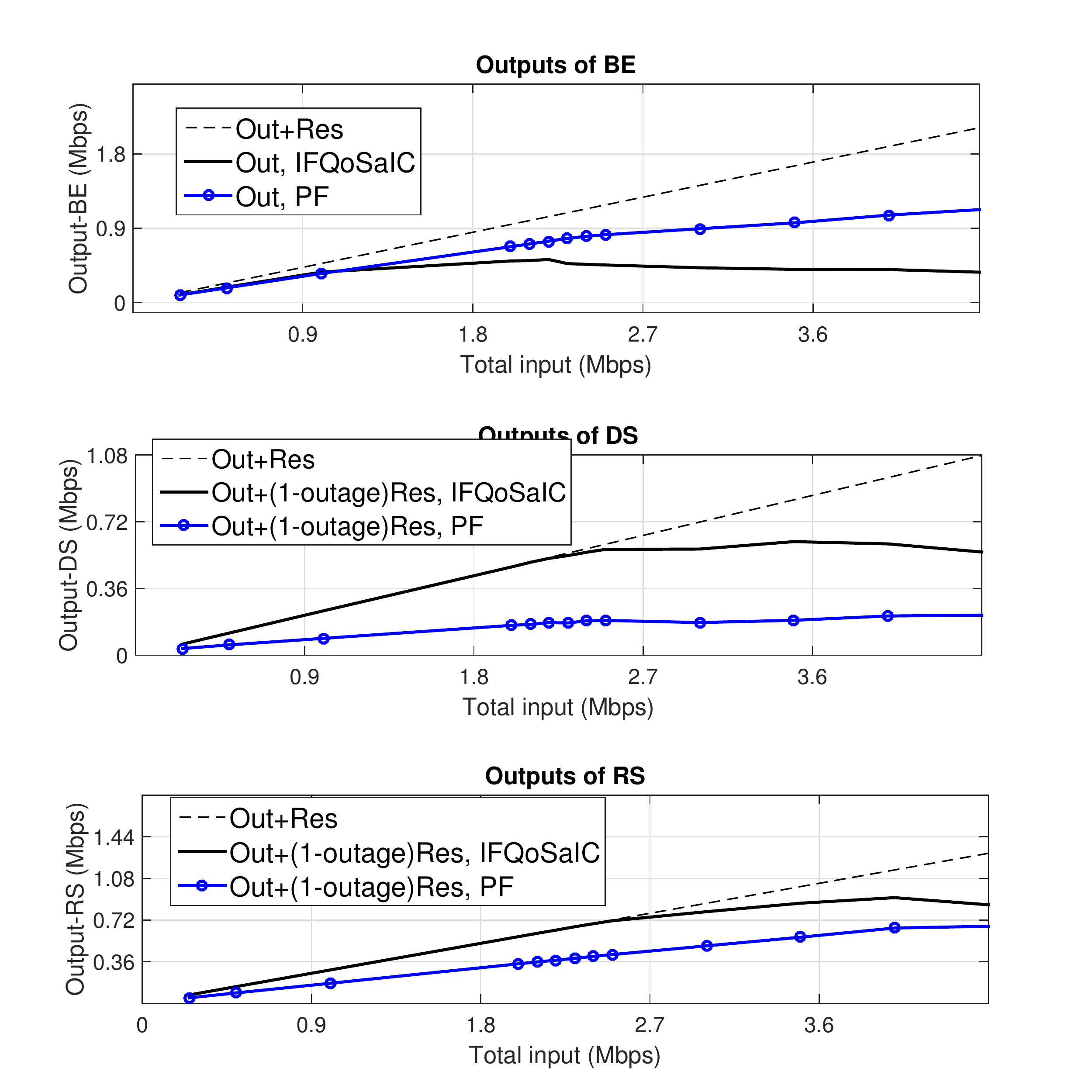}
\captionsetup{justification=centering}
\vspace{-0.4cm}
\caption{Output rates, BE+DS+RS experiment}
\label{fig:ThirdTest_ObjVsLoad}
\end{figure}

\vspace{-\BeforeSectionTitle}
\subsubsection{\underline{BE+RS+DS (LQ/HQ)}}
\label{sec:LQHQ}
\vspace{-\AfterSectionTitle}
In this experiment, we test a QoS scenario similar to the previous experiment. However, we modify the input rates to the DS flows, and separating them into two streams: one resembling a low quality (LQ) stream and one with high quality (HQ) stream.

Figures~\ref{fig:FourthTest_ObjVsOutDS} and~\ref{fig:FourthTest_ObjVsOutRS} depict the scatter plots of the objective and DS outage and objective and RS outage, respectively. We observe that for DS flows, in Figure~\ref{fig:FourthTest_ObjVsOutDS}, our algorithm outperforms the baseline in both criteria, until the $10$-th load point. Beyond that, for the overloaded conditions, our algorithm performs significantly better for DS flows (for both criteria), but has lower output for BE flows. We observe the same trend for the RS flows, in Figure~\ref{fig:FourthTest_ObjVsOutRS}.

Figure~\ref{fig:FourthTest_ObjVsLoad} plots the amended output rates versus input rates. Similar to the previous experiment, the upper, the middle, and the lower figures plot the amended output rate of BE flows, DS flows, and RS flows (separately for our algorithm and the PF). Here, when using our algorithm, the priority of sensitive flows penalizes BE flows. At this cost, we observe that the outputs of RS and DS flows are significantly higher for our algorithm for all loads. Furthermore, we observe that the output of sensitive flows saturates in overload conditions, agreeing with Figures~\ref{fig:FourthTest_ObjVsOutRS} and~\ref{fig:FourthTest_ObjVsOutDS}. We also note that, in this experiment, the BE flows suffer further (in comparison to previous experiments) due to the higher load of the sensitive flows.

The figures confirm an important characteristic of the proposed algorithm: the heterogeneous DS flows (different input rates) are handled \textit{automatically} based on the translation in~\eqref{eq:Rmin2FromDelay}, using just their delay requirements; there is \textit{no} need to set rate sensitivity for DS flows, with different arrival rates. The proof for this claim is that the algorithm is able to keep the DS outage close to zero, in Figure~\ref{fig:FourthTest_ObjVsOutDS}, without knowing DS flows input rates. This is done automatically based on the translation in \eqref{eq:Rmin2FromDelay}, where it increases required minimum rate for DS flows, when the input load increases.

Finally, in overloaded conditions, the decrease in rates of BE flows is due to fact that the constraints to the problem get tighter. Therefore, when comparing two different overloaded conditions, attempting to satisfy the constraints at the higher load \textit{reduces} the achievable value of the objective function (in comparison to the lower load). During other simulation experiments, we also observed that the artefact (especially manifested in Figure~\ref{fig:FourthTest_ObjVsOutRS}) in transiting from the close-to-zero outage to high outage is lessened with increasing the number of frames. Also, the closer to $1$ we choose the decrease parameter ($\gimel^{\mathrm{dec}}$) in \textsc{ILM}, the less the drop (in output) at the edge of underload/overload (at the cost of higher computational complexity).

\begin{figure}[t!]
\includegraphics[width=1\linewidth]{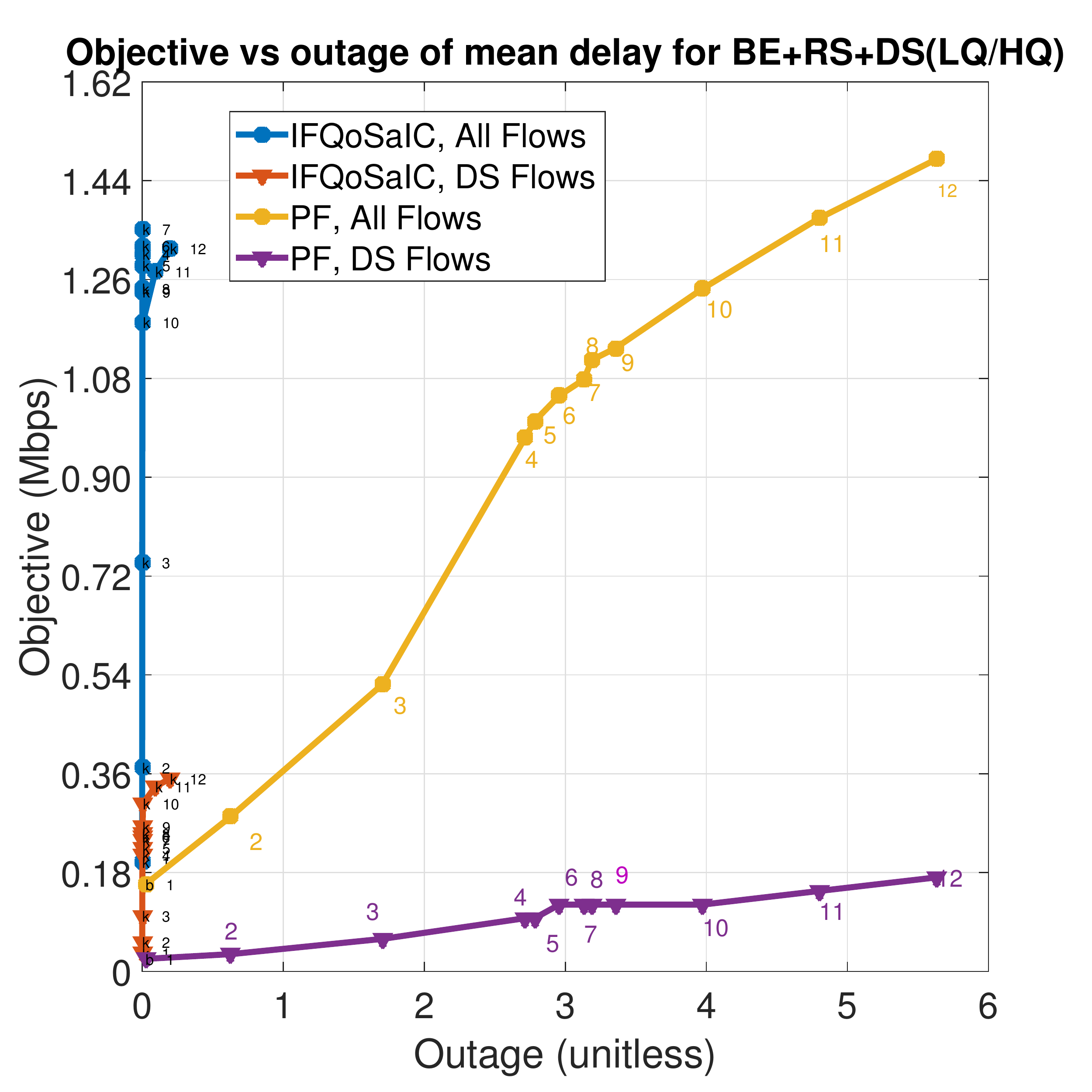}
\captionsetup{justification=centering}
\vspace{-0.4cm}
\caption{Scatter plot of objective and DS outage, BE+RS+DS (LQ/HQ) experiment}
\label{fig:FourthTest_ObjVsOutDS}
\end{figure}

\begin{figure}[th]
\includegraphics[width=1\linewidth]{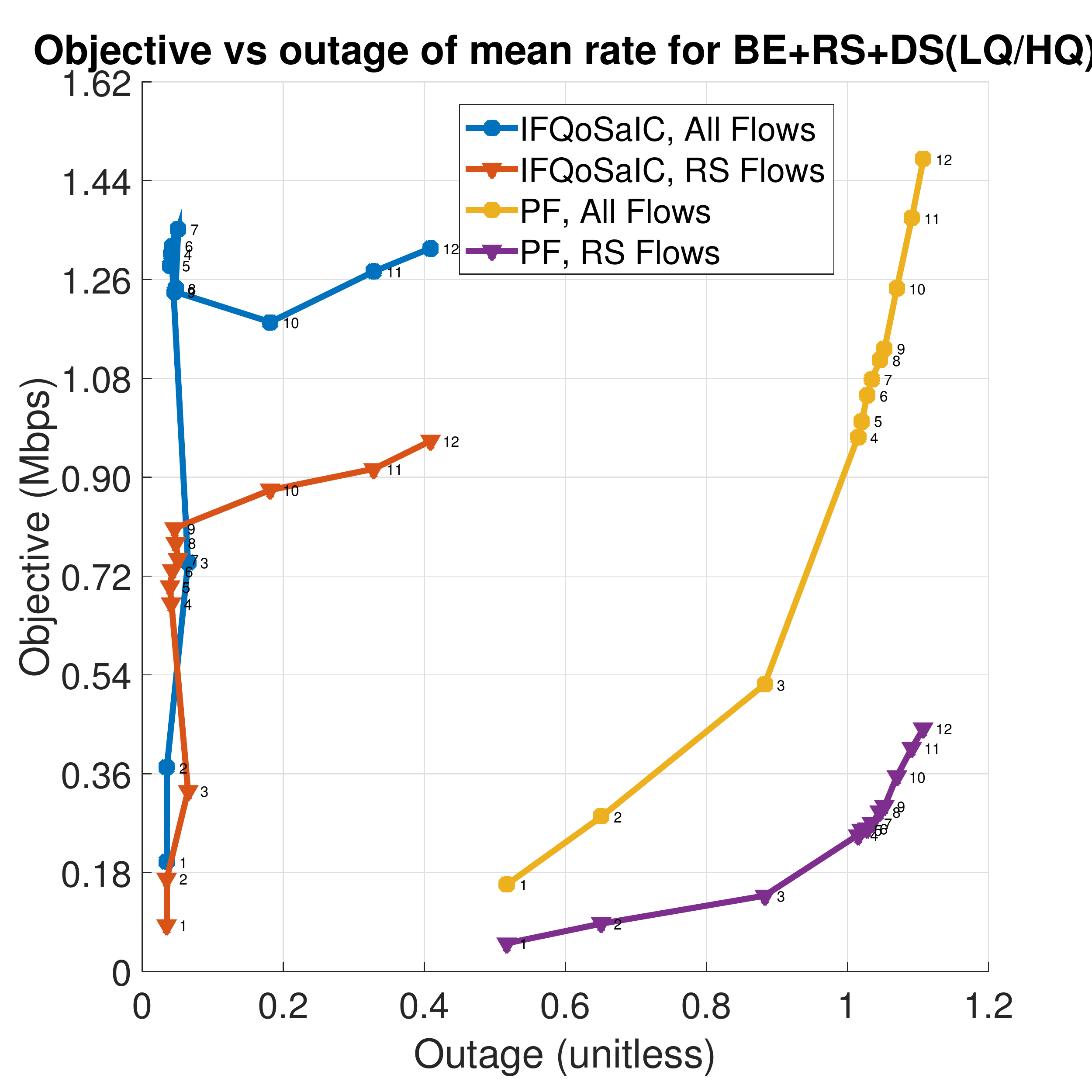}
\captionsetup{justification=centering}
\vspace{-0.4cm}
\caption{Scatter plot of objective and RS outage, BE+RS+DS (LQ/HQ) experiment}
\label{fig:FourthTest_ObjVsOutRS}
\end{figure}

\begin{figure}[th]
\centering
\includegraphics[width=1\linewidth]{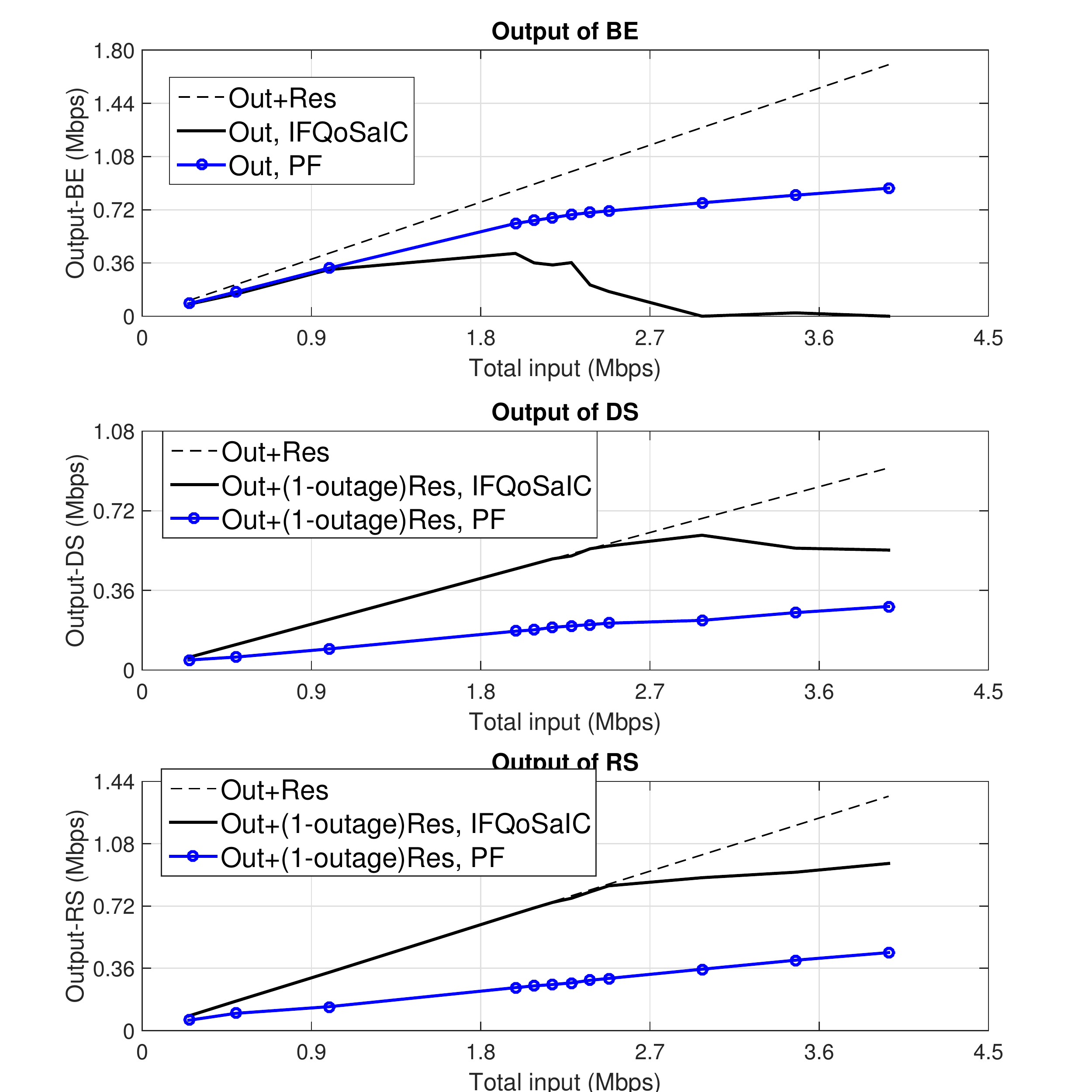}
\captionsetup{justification=centering}
\vspace{-0.3cm}
\caption{Output rates, BE+DS+RS (LQ/HQ) experiment}
\label{fig:FourthTest_ObjVsLoad}
\end{figure}

\vspace{-\BeforeSectionTitle}
\subsection{Discussion and Remarks}
\label{sec:remarks}
\vspace{-\AfterSectionTitle}
\noindent We wrap up this section with a number of important remarks:

\begin{itemize}
\item \textit{Prioritization:}
In low load conditions, outages are zero and the objective increases with the load. In this underload condition, since the available resources are more than adequate to serve the sensitive flows without outage, BE flows are served. However, in an overloaded condition (non-zero outage region), a BE flow is served only \textit{if it does not cause harmful interference to the RS/DS flows}. The user and AP location scenario (and backlog condition) determines whether a BE flow gets served or not. As a consequence of prioritizing the RS/DS flows, the \emph{total} output rate \emph{might drop}. This can be seen, for example, in Figures~\ref{fig:ThirdTest_ObjVsOutDS} and~\ref{fig:ThirdTest_ObjVsOutRS}.

Note that we do \textit{not} use any external prioritization and rely only on \textit{automatic prioritization} based on the \emph{relative} values of the translated min frame rates (generated in \textsc{QoSiFT}). We also note that the user association is \textit{not exclusively} based on channel strengths, but rather is a function of \emph{relative} channels (considering interference), traffic load, and QoS constraints of all the flows. This provides network-wide interference-, channel-, QoS-, and traffic-aware joint load balancing.

\item \textit{Behaviour of outage:}
For an RS flow, it is obvious that outage should increase when the load increases (as its input rate is increased). For a DS flow (see for example Figure~\ref{fig:FirstTest_ObjVsOutDS}), however, it is worth highlighting that although the delay requirements are fixed, increasing the input load increases the translated minimum frame rate requirements. The minimum frame rate requirement (see~\eqref{eq:MeanDelayEst} \& $\zeta^{(1)}_{\phi} [k]$) is an increasing function with respect to both $\bar{q}_{\phi} [k]$ and $q_{\phi} [k]$. Therefore, the observed increase in outage with increasing load is intuitive, despite the fixed delay targets.

\item \textit{Intra-cell interference:}
This remark is on the explicit constraint for intra-cell interference, particularly, related to the footnote on page~\pageref{sec:IntercellInterferenceExplanation}. We tested four setups of combining \emph{including/not including intra-cell interference}~\eqref{eq:IntraCellInterference} in the interference metric with \emph{including/not including explicit intra-cell interference constraint}~\eqref{eq:constRBinCellPortsVer1}. We found that we need both. Nevertheless, inspired by~\cite{ZheAdve}, we also tested eliminating the explicit intra-cell constraint, \emph{while checking for the constraint satisfaction in the termination condition} (Step~$15, 18$ of \textsc{QoSaIC}). We found by experiment that this approach reduces the computational complexity. This is because the event of violating a QoS constraint (due to eliminating the intra-cell interference) is a rare event.

\item \textit{Frugality constraint:}
	We also investigated a version of the formulation, in which we used an explicit frugality constraint. Accordingly, we tested an \textsc{ILM}, based on relaxing \textit{both} the minimum and maximum frame requirements. However, we found that an explicit $r^{\mathrm{max}}_{\phi} [k]$ constraint produces unintended consequences, including making the optimization infeasible, and triggering the \textsc{ILM}, in low load conditions. When this trigger happens, \textsc{ILM} \emph{unnecessarily} compromises the minimum rate requirement in a frame, while the minimums are feasible and infeasibility is due to the maximum rate constraints (and/or the situation that the lowest RB capacity is larger than the margin between max and min targets). This approach could not distinguish the necessity of whether to compromise minimum or maximum requirements, and led to QoS outages in low load conditions. We alleviated this issue by approximating the frugality constraint in Section~\ref{sec:SoftFrugalityConstraint}.

\item \textit{Soft constraints:}
We also tested several other methods of soft constraints for the min frame rate requirements, similar to the frugality constraint. However, none of them was as effective as the proposed \textit{explicit} constraint for the QoS requirements.
\end{itemize}

\label{AddedAnalytic}
Finally, we highlight here that in this paper, we have not considered finding the analytic performance evaluation. Analytic expressions in our work would be inherently hard to find, becuase we designed for multicell, multi QoS classes, dynamic interference coordination, and dynamic user association. It is in contrast to works, such as \cite{stolyar2001largest, AnalyticDelay} providing important analytic evaluation but for rather simple legacy designs in simple settings.

\vspace{-\BeforeSectionTitle}
\section{Conclusions}
\label{sec:conc}
\vspace{-\AfterSectionTitle}
Motivated by the need to maximize the use of the limited frequency resources and to address the growing heterogeneity in QoS demands, we formulated a framework for network-wide optimization accounting for heterogeneous QoS (DS, RS, and BE applications), dynamic interference coordination, and dynamic user association. The abstraction collects inputs of channel strengths, traffic classes, QoS requirements, input load, and makes decisions on queue/RB scheduling, dynamic user association, and dynamic interference coordination. The problem is complex, as it constitutes multiple interconnected cells; is coupled over flows and has memory over frames.

We observed that the effective network capacity changes with the AP/user scenario and QoS requirements. The RRM decisions respond not only to user location (and channels), but also to the QoS requirements and input load. Our solution exploits the opportunities in the QoS classes and the interference scenarios, in order to match capacity with demands. Our proposed suite of algorithms outperforms the baseline significantly, for both zero outage and non-zero outage regions, therefore, we conclude that our algorithms importantly have \emph{graceful degradation}.

We used tailored methods from primal-dual theory, fixed point iterations, and iterative complimentary slackness, to develop systematic solutions. In order to provide insights, we also discussed the interplay between the parameters in derivations and algorithms. In the simulation, we observed the significant gains made possible by our algorithms. Moreover, we made important observations, summarized in Section~\ref{sec:remarks}, about the general trends of output rates and outages versus input rates, rate saturations, automatic flow prioritization, effects in the transition from underload to overload, handling heterogeneous DS flows without knowing their input rates, dealing with the frugality constraint, and in which load situations, BE flows can be served.

In summary, this paper provides framework and solution crucially for heterogeneous QoS \textit{jointly} with \textit{dynamic interference coordination and user association} exploiting the joint processing made possible by C-RAN. The result of this joint processing increases the effective capacity of the system, enhances users' experience, and enables provisioning new services (especially, centered around delay constrained applications).

\section*{Acknowledgement}
The authors would like to acknowledge Dr. Ivo Maljevic and Dr. Halim Yanikomeroglu for their support and comments. We also appreciate the associate editor and anonymous reviewers for their valuable comments and suggestions which improved the presentation and the content of the paper.

\bibliographystyle{IEEEtran}
{
\bibliography{UniBIB}}

\vspace{-1cm}
\begin{IEEEbiography}
[{\includegraphics[width=1.5in,height=1.5in,clip,keepaspectratio]{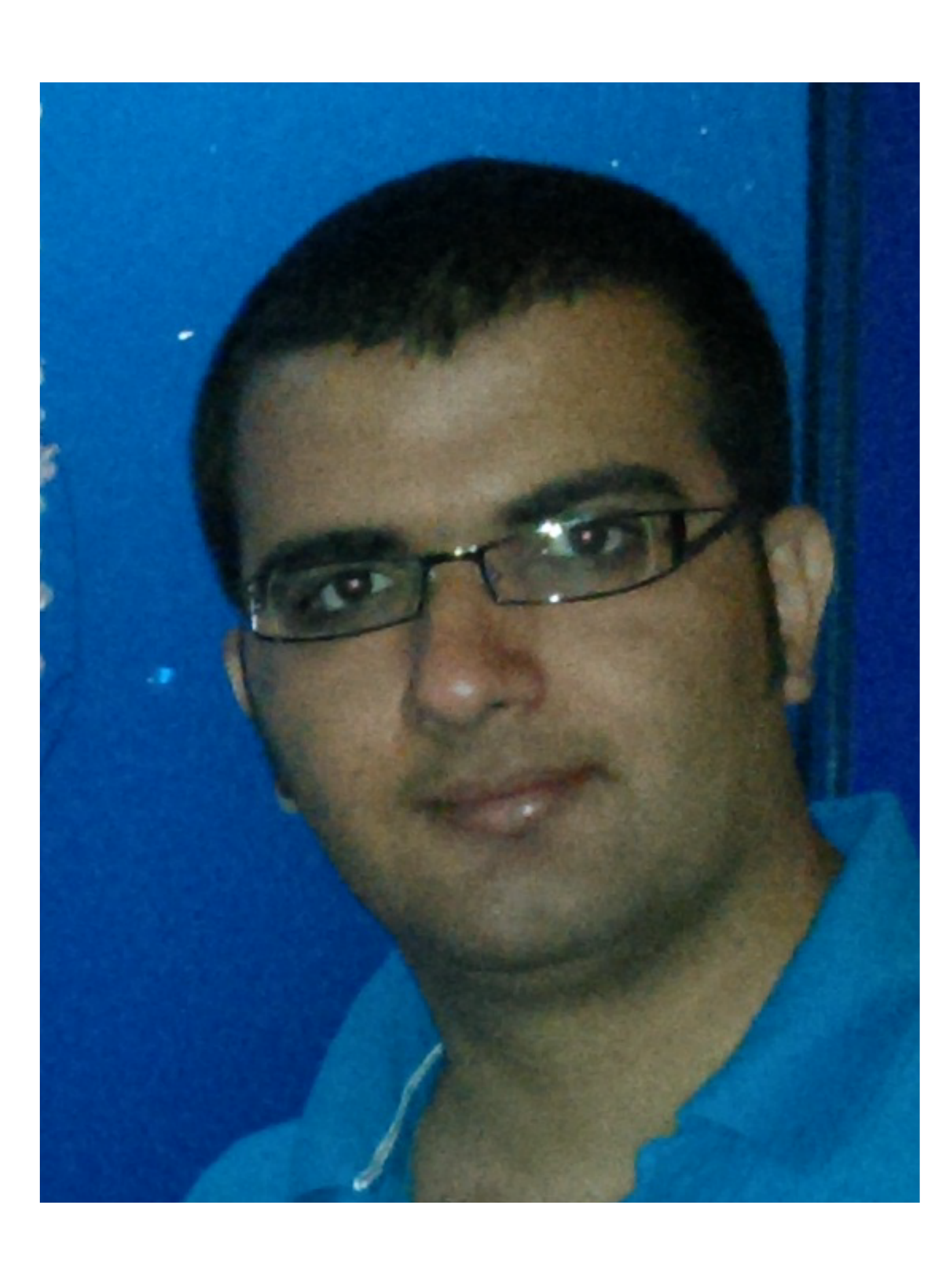}}]{Alireza Sharifian} (M'14) was born in Isfahan, Iran. He received the B.Sc. degree in Electronics (2005), the M.Sc degree in Communications (2008), from Isfahan University of Technology, Isfahan, Iran, and the Ph.D. in Electrical and Computer Engineering (2014), from Carleton University, Ottawa, Canada, with Senate medal distinction for outstanding academic achievement. Throughout his PhD, he was a member of a strategic project with Huawei-Shenzhen and Huawei-Ottawa R\&D offices, supported by ORF-RE (Ontario Research Fund - Research Excellence). He is currently a research fellow with the Edwards S. Rogers Sr. Department of Electrical and Computer Engineering, University of Toronto. Dr. Sharifian was a recipient of TelentEdge Ontario Centers of Excellence. His research interests, in 5G, include Cloud-RAN, HetNets, cooperative multipoint systems, resource management, scheduling, QoS, relay networks, OFDMA, IoT, and LTE standards. His other line of interests and expertise comprise signal processing, data science, machine learning, reinforcement learning, dimensionality reduction, graph Laplacian applications, and optimization. He also served as reviewer for several publications, including IEEE Trans. on Wireless Comm., IEEE Tran. on Vehicular Tech., and IEEE Trans. on Comm.
\end{IEEEbiography}

\vspace{-1cm}
\begin{IEEEbiography}
[{\includegraphics[width=1.35in,height=1.35in,clip,keepaspectratio]{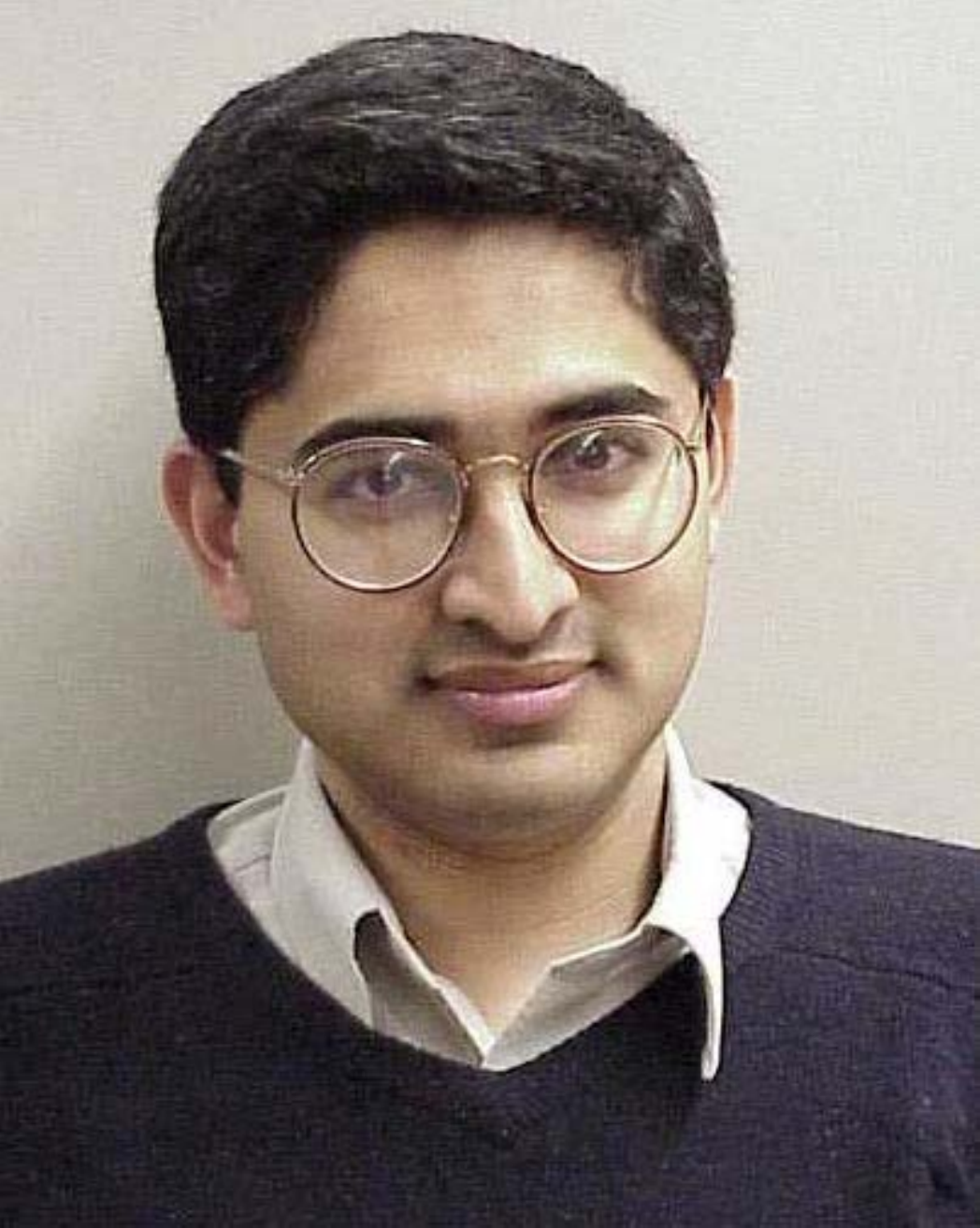}}]
{Raviraj S. Adve} (S'88-M'97-SM'06-F'17) was born in Mumbai, India. He received the B.Tech. degree in electrical engineering from IIT Bombay, Mumbai, in 1990, and the Ph.D. degree from Syracuse University in 1996. From 1997 to 2000, he was with Research Associates for Defense Conversion Inc., on contract with the Air Force Research Laboratory, Rome, NY, USA. He joined as a Faculty Member with the University of Toronto in 2000, where he is currently a Professor. His research interests include practical signal processing algorithms for multiple input multiple output, wireless communications, and distributed radar systems. In the area of wireless communications, he is currently focused on cooperation in distributed wireless networks. In radar systems, he is particularly interested in waveform diversity and low-complexity space-time adaptive processing algorithms. He received the 2009 Fred Nathanson Young Radar Engineer of the Year award.
\end{IEEEbiography}

\end{document}